# Unveiling Zn incorporation in CuInS$_2$ quantum dots: X-ray and optical analysis of doping effects, structural modifications and surface passivation

Andrés Burgos-Caminal[1,2*], Brener R. C. Vale[3,4], André F. V. Fonseca[6,4], Juan F. Hidalgo[1], Elisa P. P. Collet[1], Lázaro García,[2] Víctor Vega-Mayoral[1], Saül Garcia-Orrit[1], Iciar Arnay[1], Juan Cabanillas-González[1], Laura Simonelli[5], Ana Flávia Nogueira[6], Marco Antônio Schiavon[4], Thomas J. Penfold[7], Lazaro A. Padilha[3] and Wojciech Gawelda[2,1,8*]

*1. Madrid Institute for Advanced Studies IMDEA Nanoscience, Ciudad Universitaria de Cantoblanco, Calle Faraday 9, 28049 Madrid, Spain*
*2. Departamento de Química, Universidad Autónoma de Madrid, Ciudad Universitaria de Cantoblanco, Calle Francisco Tomás y Valiente 7, 28049 Madrid, Spain*
*3. Ultrafast laboratory spectroscopy, "Gleb Wataghin" Institute of Physics, University of Campinas, Brazil*
*4. Grupo de Pesquisa Química de Materiais, Departamento de Ciências Naturais, Universidade Federal de São João Del-Rei, Brazil*
*5. CELLS-ALBA Synchrotron Light Source, 08290 Cerdanyola del Vallès, Barcelona, Spain*
*6. Laboratório de Nanotecnologia e Energia Solar, Chemistry Institute, University of Campinas – UNICAMP, Campinas, São Paulo, Brazil*
*7. Chemistry, School of Natural and Environmental Sciences Newcastle University, NE1 7RU Newcastle upon Tyne, UK*
*8. Faculty of Physics, Adam Mickiewicz University, ul. Uniwersytetu Poznańskiego 2, 61-614 Poznań, Poland*

*Corresponding author(s): andres.burgos@imdea.org ; wojciech.gawelda@uam.es*

**Abstract:**

Quantum dots (QDs) exhibit unique properties arising from their reduced size and quantum confinement effects, including exceptionally bright and tunable photoluminescence. Among these, CuInS$_2$ QDs have gained significant attention owing to their remarkable broadband emission, making them highly desirable for various optoelectronic applications requiring efficient luminescent nanomaterials. However, maximizing radiative recombination in CuInS$_2$ QDs often necessitates minimizing intragap trap states. A common approach involves the introduction of Zn during the synthesis, which typically promotes the formation of a ZnS shell that passivates the QD surface.

Despite its importance, the characterization and quantification of Zn incorporation using conventional techniques, such as optical spectroscopy or electron microscopy, remains challenging. In this study, we utilized X-ray absorption spectroscopy (XAS), in both X-ray absorption near-edge structure (XANES) and extended X-ray absorption fine structure (EXAFS) spectral ranges, to investigate Zn incorporation into CuInS$_2$ QDs with element-specific precision. This approach allowed us to detect the formation of a ZnS surface shell and to resolve the spatial distribution of Zn atoms within the QD lattice, distinguishing between Zn as a substituent, or as an interstitial defect.

Additionally, we explored the optical and dynamical properties of CuInS$_2$ QDs using time-resolved optical spectroscopies, particularly in the presence of electron and hole acceptors. These results provide deeper insights into the role and effectiveness of the Zn-induced passivating layer, paving the way for optimizing QD performance in photoluminescence applications.

**Keywords:** CuInS$_2$ quantum dots, X-ray absorption, Zn doping, Core-shell, charge carrier dynamics, charge transfer

## Introduction

CuInS$_2$ quantum dots (CIS QDs) have been intensively studied for different technological applications,[1] such as light emitting diodes,[2,3] luminescent solar concentrators,[4–6] QD sensitized solar cells,[7,8] fluorescent probes,[9,10] or cell imaging.[11,12] One of the most important aspects in many of these applications, especially in optoelectronics, is the photoluminescence (PL) quantum yield, which is given by the ratio between emitted and absorbed photons.[13] Photons are generally emitted through the radiative recombination of excitons inside the QD. Any other process that involves exciton recombination without the emission of a photon will be detrimental and decrease the PL quantum yield (PLQY). Typically, at low excitation fluences, non-radiative recombination is dominated by trap-assisted recombination akin to the Shockley-Read-Hall mechanism in bulk semiconductors.[14] Although the defect states involved in this mechanism can be present inside the core of the QD, they are most often localized on the surface, due to the unbound orbitals of the surface atoms. A common strategy to passivate surface defects is to employ a shell of a larger bandgap semiconductor material surrounding the core of the QD, to prevent the formation of surface trap states, and generally confine the photoexcited charge carries inside the core.[15,16]

CIS QDs are typically passivated with a ZnS shell, although other structures such as CuInS$_2$/CdS, have also been reported.[17] ZnS is particularly favorable due to the similar structure of ZnS zinc blende and CIS chalcopyrite. A progressive blue shift of the emission with the formation of the shell has been interpreted as ZnS forming an alloy with a composition gradient across the interface instead of a clear and distinct shell.[18] Indeed, according to Berends *et al.*,[19] the reaction growth of ZnS with CIS QDs is more complex than that of other QDs, and this reaction can result in alloy formation, cation exchange, etching, and core-shell formation. Furthermore, Zn$^{2+}$ cations are often considered to replace Cu$^+$ and In$^{3+}$. This can occur when applying a post-treatment with a Zn$^{2+}$ precursor on previously synthesized CIS QDs, which gradually forms the ZnS shell.[18,20,21]

Usually, these different surface reactions are followed by a) indirect measurements such as UV-Vis absorption, and emission spectroscopies, or b) more direct and structure-sensitive, techniques such as X-ray diffraction, X-ray photoelectron spectroscopy, and transmission electron microscopy (TEM). However, these latter methods can only be used with solid samples, whereas optical probes can also be used in colloidal dispersion. Measuring physico-chemical properties in colloidal suspension is advantageous because the properties of QDs in solid state may be affected by the different dielectric environment.[22] Despite their inherent limitations, PL and UV-Vis) absorption spectroscopy remain the most widely utilized techniques for probing the optical properties and electronic transitions in QDs. However, the availability of more advanced and selective characterization methods—particularly those capable of analyzing nanocrystals within colloidal dispersions—is critically important for gaining deeper insights into their structural, compositional, and dynamic properties. Here, we employed steady-state X-ray absorption spectroscopy (XAS) to investigate colloidal CIS QDs. We exploit several unique advantages of XAS, such as elemental specificity, sensitivity to oxidation states and local atomic structure, compared to UV-Vis spectroscopy.[23] To maximize the obtained information, we probed both the X-ray absorption near edge structure (XANES), and the extended X-ray absorption fine structure (EXAFS) regions.[23]

We have studied five different samples, by varying their stoichiometries and Zn-doping levels, to systematically characterize and corroborate the optical properties of the material. Our main results deliver new insights into the role and incorporation of Zn atoms into CIS QD structures. We observed them both at the surface and inside the core, depending on the Cu:In stoichiometry. We probed these effects using complementary steady-state UV-Vis and X-ray spectra as relevant observables. Finally, we correlated these structural and compositional changes in CIS QDs with the charge transfer efficiency towards electron and hole acceptors using time-resolved and steady-state optical spectroscopies.

**Experimental Section**

The following methods and setups have been published previously in Refs[24,25]. The descriptions included below were taken from these two articles.

**CIS / Core Synthesis**

First, 0.0587 g of $InCl_3$ and 0.0238 g (1:1 Cu:In; 100%) or 0.004 g (0.2:1 Cu:In; 20%) of CuCl were weighed and placed in a 50 mL 3-neck flask. To this flask, 8 mL of 1-octadecene (ODE), 60 μL of oleic acid (OA), and 250 μL of dodecanethiol (DDT) were also added. The mixture was dried under vacuum at 90°C for 30 min. During this 30 min, a mixture with 0.038 g of sulfur (S) in 3 mL of oleylamine (OAm) was taken to ultrasound for 5 min. After this time, the solution of precursors in ODE was heated to 180 °C under an argon atmosphere for 5 min. Then, the temperature was lowered to 160 °C and 2 mL of the S-OAm solution were injected into the flask, monitoring for 10 min. The solution was cooled in an ice bath to room temperature (25 °C), under stirring and in an inert atmosphere (Argon).

After the synthesis step, the suspension was transferred to a Falcon centrifuge tube and 8.0 mL of isopropanol were added to purify the NCs. The tube was then taken to the centrifuge for 10 minutes at 7000 rpm. Finally, the supernatant was removed, and the nanoparticles were suspended in cyclohexane.

**CZIS / Core Synthesis**

First, 0.0587 g of $InCl_3$, 0.0238 g (1:1 Cu:In; 100%) or 0.004 g (0.2:1 Cu:In; 20%) of CuCl, and 0.0274 g of $ZnCl_2$ were weighed and placed in a 50 mL 3-neck flask. To this flask, 8 mL of 1-octadecene (ODE), 60 μL of oleic acid (OA) and 250 μL of dodecanethiol (DDT) were also added. The mixture was dried under vacuum at 90°C for 30 min. During this 30min, a mixture of 0.0257 g of sulfur (S) in 2 mL of oleylamine (OAm) was taken to ultrasound for 5 min. After this time, the solution in the 3-neck flask was heated to 180 °C under an argon atmosphere for 5 min. Then, the temperature was adjusted to 160 °C, and 2 mL of the S-OAm solution was injected into the flask, allowing it to react for 10 min. The solution was cooled in an ice bath to room temperature (25 °C), under stirring and in an inert atmosphere (Argon).

After the synthesis step, the suspension was transferred to a Falcon centrifuge tube and 8.0 mL of isopropanol were added to purify the NCs. The tube was then taken to the centrifuge for 10 minutes at 7000 rpm. Finally, the supernatant was removed, and the nanoparticles were suspended in cyclohexane.

**CZIS Core-Shell Synthesis:**

**Synthesis of Zn-OAm stock solution:**

First, 0.2725 g of $ZnCl_2$ were weighed and placed in a 50 mL 3-neck flask. To this flask, 4 ml of octadecene (ODE) and 1 ml of oleylamine (OAm) were also added. The mixture was dried under vacuum at 90 °C for 30 min. After this time, the solution was heated at 150 °C under an argon atmosphere for 10 min. Then, the temperature was adjusted to 50 °C.

**CZIS/ZnS Core-Shell Synthesis:**

The same procedure described above for the CZIS (item 2) was done, except for the purification step. With the pristine solution at room temperature and the Zn-OAm stock solution (item 3.1) at 50 °C, 5 mL of Zn-OAm solution were injected. Then, the system was heated to 200 °C and allowed to react for 30 min. Then, the solution was cooled in an ice bath to room temperature (25 °C), under stirring in an inert atmosphere.

After that, Isopropanol was added in a 1:1 ratio to the nanocrystal suspension and centrifuged at 9000 rpm for 10 minutes. The supernatant was discarded, and the tube remained open for ~5 minutes to dry isopropanol residues. The precipitate was suspended in cyclohexane.

**Transient absorption spectroscopy**

Transient absorption measurements were conducted using a Clark-MXR CPA-1 regenerative amplifier. The fundamental of the laser (775 nm, 1kHz, 120 fs, 1 mJ) was divided into two paths. One beam supplied a non-colinear optical parametric amplifier (NOPA) to generate 520 nm pulses, and filtered to the desired fluence to pump the sample. The second beam was sent through a $CaF_2$ crystal to generate a broadband supercontinuum by self-phase modulation spanning between 380 and 720 nm which was used as the probe. Due to technical circumstances, the $Cu_{0.3}InS_2$ sample had to be probed with a supercontinuum generated with a sapphire crystal, limiting its bandwidth to 480-700 nm. A delay line was used to control the temporal delay between both pulses, which spatially overlapped on the sample. The probe pulse was divided before the sample position between a reference and a signal beam. The latter is sent through the sample, and both are collected into a prism spectrometer (Entwicklungsburo Stresing GmbH with a double CCD array. A home-made software recorded the normalized change in absorption (ΔA) in a shot-to-shot configuration. All measurements were performed at magic angle (54.7º) between the pump and probe to avoid anisotropy effects. The samples were measured in 2 mm thin quartz cuvettes with constant stirring with a magnetic bar perpendicular to the incident beam.

**Steady-state X-ray absorption spectroscopy**

The steady-state XAS spectra were obtained at the BL-22 CLÆSS Beamline from the ALBA synchrotron in Barcelona (Spain).[26] The X-ray beam is obtained from a multi-pole wiggler, and monochromatized with a double-crystal monochromator employing Si(111) crystals. The beam is focused down to a spot of 200x50 μm$^2$ at the sample position. The samples are contained in closed liquid cells with Kapton® windows, and the absorption is measured in total fluorescent yield mode.

**Time-Resolved Photoluminescence**

TRPL decay dynamics were obtained through time-correlated single photon counting (TCSPC) using a TimeHarp Picoquant multichannel time correlator. A 355 nm Nd:YAG pulsed laser from Teem Photonics was used as excitation at 1kHz. PL detection was carried out at a single wavelength using a thermoelectrically cooled Hamamatsu photomultiplier coupled to a 0.5 m spectrometer (SP2500 Princeton Instruments, Acton Research) equipped with 600 lines mm$^{-1}$ grating.

**Steady-state Absorption and Photoluminescence**

UV–vis absorption spectra were recorded using a Varian Cary 5000 UV–vis-NIR spectrophotometer (Agilent), while the photoluminescence spectra were recorded with a Hamamatsu PLQY spectrometer model C13534-11. All spectra were acquired at room temperature and in ambient conditions.

**Results and discussion**

In order to compare the photophysical properties of different synthesis stoichiometries and Zn doping, we prepared five different samples (Fig. 1.A): A) stoichiometric $CuInS_2$, B) Cu-deficient $CuInS_2$, C) Zn-doped stoichiometric $CuInS_2$, D) Zn-doped Cu-deficient $CuInS_2$, and E) Zn-doped Cu-deficient $CuInS_2$ with a post-treatment of $ZnCl_2$ at 200 ºC. This last post-treatment is expected to further passivate the surface and thus prevent non-radiative recombination and charge transfer to surface acceptors.[21] The detailed synthesis method, which is based on previously published work, can be found in the Methods section. Elemental analysis reveals that the synthesis under stoichiometric conditions produces Cu-rich samples (Table S1). However, Zn-incorporation results in a Cu:In stoichiometry closer to the nominal value from the synthesis. Considering these results we have named the samples accordingly (A) $CuIn_{0.4}S_2$, B) $Cu_{0.3}InS_2$, C) $Cu(Zn)In_{0.8}S_2$, D) $Cu_{0.3}(Zn)InS_2$, and E) $Cu_{0.2}(Zn)InS_2/ZnS$). The stoichiometry of Zn is not included in the formula due to the uncertainty of its incorporation at this point. High-resolution transmission electron microscopy (HR-TEM) measurements show sizes that range between 3.3 ± 0.5 nm for $Cu_{0.3}(Zn)InS_2$ to 5.3 ± 0.6 nm for $CuIn_{0.4}S_2$ (Fig. S1). XRD results (Fig. S2) show good agreement with a chalcopyrite structure for the Cu-deficient samples, although for Cu-rich samples we observe an additional peak at at 2θ ≅ 50º which we attribute to a partial contribution of a wurtzite structure. More precise structure determination is prevented by the increased widths of the peaks.

Fig. 1 also shows the reported crystalline structures of the $CuInS_2$ core and the ZnS shell (Fig. 1.B), as well as the UV-Vis absorption (Fig. 1.C) and PL (Fig. 1.D) spectra of the investigated samples. The absorption spectra show that Cu-deficient samples have sharper features and are blue-shifted compared to the Cu-rich ones. Zn-doped samples show further blue-shifted spectra compared to the ones without Zn. According to previous studies, this behavior is quite common for CIS QDs and is due to cation-exchange processes, in which $Cu^+$ and $In^{3+}$ are replaced by $Zn^{2+}$ in the lattice structure.[20,21] Density of state calculations have demonstrated that, for CIS QDs, $Cu^+$ and $S^{2-}$ contribute to the valence band (VB) edge, whereas $In^{3+}$ dominates the conduction band (CB) edge, with a small contribution of $Cu^+$ and $S^{2-}$.[27] However, once $Zn^{2+}$ ions are present inside the lattice, forming an alloy, $Zn^{2+}$ (3$d$) orbitals start to contribute to the CB-edge, which shifts it to higher energies with the increasing concentration of $Zn^{2+}$ in the lattice.[27] Lastly, only the $Cu_{0.2}(Zn)InS_2/ZnS$ and $Cu_{0.3}(Zn)InS_2$ samples show significant photoluminescent properties among all the samples. By looking closer at the PL signals of all the samples in the log scale (Fig. 1.D) we can observe that the weak PL band of $Cu_{0.3}InS_2$ is slightly higher than the non-detectable ones for $CuIn_{0.4}S_2$ or $Cu(Zn)In_{0.8}S_2$. This indicates that both Cu deficiency and Zn doping are beneficial for reducing charge carrier trapping.

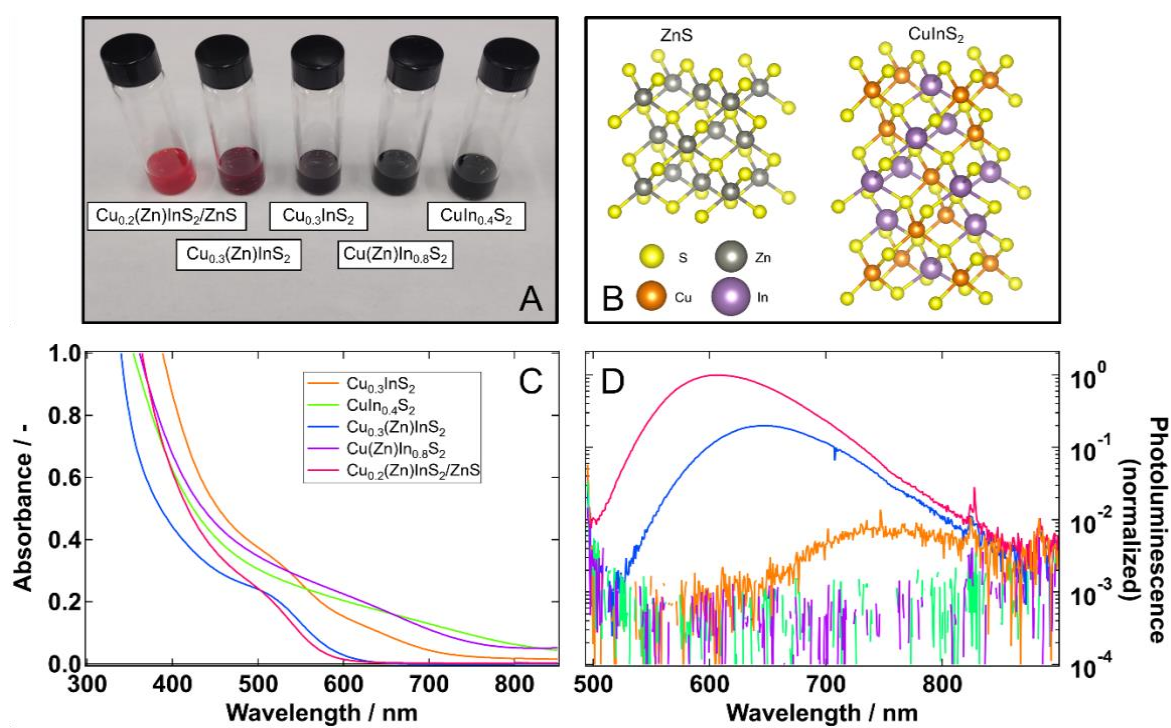

**Figure 1**: A) Photograph of the five samples under study, and B) expected crystalline structures of the $CuInS_2$ core (chalcopyrite) and the ZnS shell (zincblende).[28–30] The bottom panels show the steady state characterization of the $CuInS_2$ samples $Cu_{0.2}(Zn)InS_2/ZnS$ (red), $Cu_{0.3}(Zn)InS_2$ (blue), $Cu(Zn)In_{0.8}S_2$ (purple), $Cu_{0.3}InS_2$ (orange), and $CuIn_{0.4}S_2$ (green). C) UV-Vis absorbance, and D) log scale PL spectra of the studied samples with different levels of passivation. Note the correlation between the disappearance of the sub-bandgap tail in the absorption spectra (C) and the increase in the PL signals (D). The linear version of panel D is shown in Fig. S3A.

Furthermore, the steady state characterization in Fig. 1 shows that a considerable PL signal appears only with both Zn doping and Cu deficiency. This also eliminates most of the considerable sub-bandgap absorption tails,[31] which are assigned to defects and disorder in the CIS QDs. We experimentally obtained PLQY values of 30%, 7.4%, and 0.4% for $Cu_{0.2}(Zn)InS_2/ZnS$, $Cu_{0.3}(Zn)InS_2$, and $Cu_{0.3}InS_2$, respectively, in line with the passivation level and the PL spectra. On the other hand, both Cu-rich samples exhibit what appears to be a localized surface plasmon resonance (LSPR) in the near-infrared (NIR), as shown in Fig. S3.B, which is an unexpected result. The occurrence of LSPR has been reported for samples with Cu-deficiency, which is reported to lead to hole doping and eventually to the formation of surface plasmons.[32] Our results suggest it could likewise be caused by the Cu-excess in CIS QDs, providing electron doping.

**Structural investigation**

In order to shed more light on the Zn doping effect and the details of its spatial distribution within the crystal lattices of the different QD structures, we carried out a systematic structural analysis performing XAS experiments at the CLÆSS beamline of the ALBA synchrotron (Barcelona, Spain).[26] Our aim was to correlate XAS signals originating from different atomic constituents of the QDs, i.e. from Zn, Cu and S atoms, which can be found both inside and on the surface of the nanocrystals. Given the very broad energy range of CLÆSS (2.4-63 keV) and its versatile equipment for measuring XAS signals for both solid and liquid samples, we were able to record XANES and EXAFS spectra for colloidal dispersions of CIS QDs at Zn, S and Cu K-edges in a single experiment. The most relevant for this study are the XAS signals for the Zn and S atoms (Cu is not shown here, except in Fig. S4). The results for the Zn K-edge are shown in Fig. 2, where Fig. 2.A depicts the normalized XANES spectra of the different $Cu(Zn)InS_2$ (CZIS) samples studied and the corresponding signal for a ZnS reference. Fig. 2.B shows the Fourier transform of the corresponding $k^2$-weighted EXAFS oscillations.

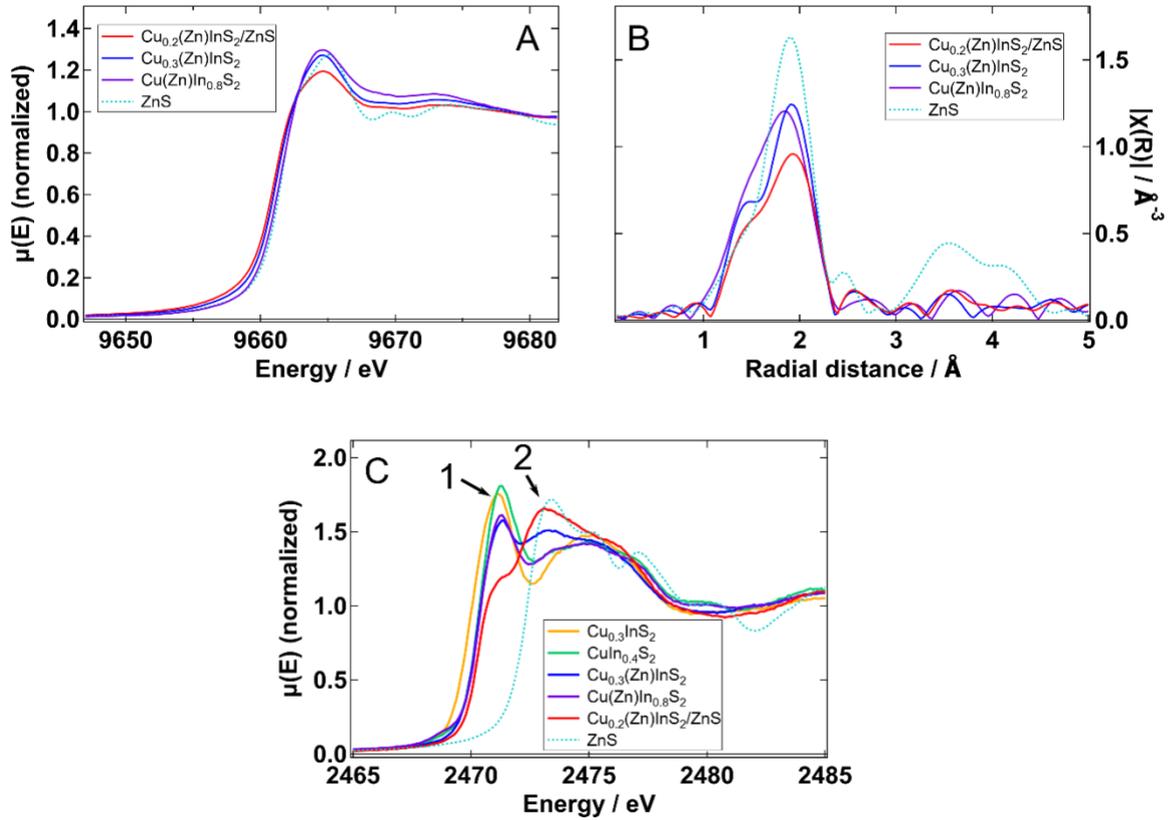

**Figure 2:** A) Normalized XANES spectra at the Zn K-edge of the Zn-containing samples and bulk ZnS, and B) the Fourier transforms of the corresponding $k^2$-weighted EXAFS oscillations. C) Normalized XANES spectra at the S K-edge. Two arrows point to the characteristic bands originating from S atoms inside the QD (**1**) and within the ZnS shell on the surface (**2**).

For all CZIS QDs, the Fourier-transformed EXAFS signals contain not only the main peak, which originates from the scattering of the nearby sulfur atoms (first coordination shell at around 1.9 Å), but also a non-negligible and reproducible shoulder at lower distances. This can be assigned to either impurities that bond to Zn with shorter bond lengths,[33] or to Zn at interstitial or displaced positions.[34,35] Although interstitials contributions would be expected to be detected at around 1 Å, it is hard to univocally identify them because of a combination of noise level and background subtraction which can induce artifacts. Instead, the visualization of the corresponding wavelet transform[36] can allow us to access finer details (see Fig. 3, further below in the text).

Interestingly, comparing the three Zn-doped QDs, a larger difference between the Cu-rich (Cu(Zn)In$_{0.8}$S$_2$) and the Cu-deficient samples (Cu$_{0.3}$(Zn)InS$_2$ and Cu$_{0.2}$(Zn)InS$_2$/ZnS) can be observed. For the Cu-rich sample, the main peak shifts to lower distances and presents a more pronounced low-distance shoulder at the same time. This points to a different incorporation of Zn for the two cases. While for Cu(Zn)In$_{0.8}$S$_2$, Zn atoms are incorporated mainly inside the QDs, the other two cases (Cu$_{0.3}$(Zn)InS$_2$ and Cu$_{0.2}$(Zn)InS$_2$/ZnS) have a large proportion of Zn atoms at the surface, probably forming a shell of ZnS. Qualitatively, this interpretation is further corroborated by comparing the positions of the main Fourier transform features for the Cu-deficient QDs and the ZnS bulk sample (Fig. 2.B).

In order to access finer details, we have performed EXAFS fitting using FEFF simulations carried out with the Demeter package.[37] In our model, we used a zincblende crystal structure of ZnS[30] for fitting the first coordination shell only (Fig. S5 and Table S2). The main output extracted from this analysis is the coordination number, which is ideally equal to 4. For the $Cu(Zn)In_{0.8}S_2$ sample, we obtained a fit result close to 4, however, for $Cu_{0.3}(Zn)InS_2$ this number decreased to 3.5, and for $Cu_{0.2}(Zn)InS_2/ZnS$, it decreased even further down to 2.6. This decrease in coordination number indicates an increased proportion of Zn at the QDs surface. The quality of our fits indicates that the actual structure surrounding Zn atoms is much more complex than the bulk structures used for the fit. Nonetheless, they can serve for a semi-quantitative analysis of changes in the local coordination of Zn atoms.

Looking at the XANES spectra (Fig. 2.A), the three QD samples present a peak centered at 9664.5 eV. It corresponds to the dipole-allowed $1s \rightarrow 4p$ transition and its position exhibits energy shifts, which can be correlated with the coordination.[38] We observe that it broadens following $Cu(Zn)In_{0.8}S_2$ < $Cu_{0.3}(Zn)InS_2$ < $Cu_{0.2}(Zn)InS_2/ZnS$. This agrees with what is observed in the EXAFS region, and it denotes that as we increase the passivation, the fraction of Zn atoms present on the QDs surface increases. Contrary to Zn atoms localized inside the core, those on the surface are characterized by higher disorder and lower coordination numbers (see EXAFS results mentioned earlier).

We now turn our attention to sulfur atoms and discuss the S K-edge XANES spectra shown in Fig. 2.C. For all studied samples we observe a main peak at 2471.3 eV (peak 1) which corresponds to the $1s \rightarrow 3p$ dipolar transition.[39,40] Filling the $3p$ orbitals of sulfur anions by ionic bonding decreases the intensity of this peak. For instance, S with a formal charge of -2 should exhibit a very low-intensity peak. Fig. 2.C shows that peak 1 decreases in the following order: $CuIn_{0.4}S_2$ > $Cu_{0.3}InS_2$ > $Cu(Zn)In_{0.8}S_2$ > $Cu_{0.3}(Zn)InS_2$ > $Cu_{0.2}(Zn)InS_2/ZnS$.

These observations can be explained in light of similar findings in other semiconductors, such as $Cu_xIn_ySe_2$ (CISe), with a chalcopyrite crystal structure, which behaves similarly to CIS.[39,41] Yamazoe *et al.* showed that Se K-edge peak 1 intensities are related to the relative amounts of copper and indium in $CuInSe_2$ structures and the Se coordination environment. CISe with a 3-fold coordinated Se and more Cu vacancies show a higher intensity of peak 1 than that in CISe with a 4-fold coordinated Se and fewer Cu vacancies.[41] Because S behaves in a similar way to Se, we can interpret the increase in intensity of peak 1 in $CuIn_{0.4}S_2$ and $Cu(Zn)In_{0.8}S_2$, as compared to $Cu_{0.3}InS_2$ and $Cu_{0.3}(Zn)InS_2$, as an indication of a more distorted tetrahedral S environment for the Cu-rich samples. This conclusion is in line with what has been observed for the Cu K-edge in our previous publication[24] and in Fig. S4. Yamazoe *et al.* also discusses the appearance of an additional peak,[41] which in our case is present at 2475 eV. However, we do not include it in the discussion because it overlaps with other spectroscopic features. Alternatively, peak 2 (2473.1 eV), which is present and well defined for samples $Cu_{0.3}(Zn)InS_2$ and $Cu_{0.2}(Zn)InS_2/ZnS$, corresponds to S in a ZnS environment,[40,42] agreeing with the bulk ZnS spectrum. Therefore, this result corroborates that ZnS shells are formed at the QD surface for these two samples, with $Cu_{0.2}(Zn)InS_2/ZnS$ having a considerably larger thickness, also in agreement with the result of the Zn K-edge.

This also provides direct evidence to explain their considerable PL signal strength due to increased surface passivation. Its presence in Cu(Zn)In$_{0.8}$S$_2$ is, however, very limited and difficult to confirm.

To further quantify the difference between Cu$_{0.3}$(Zn)InS$_2$ and Cu$_{0.2}$(Zn)InS$_2$/ZnS, a linear combination fit of the first one was carried out as shown in Fig. S6. An acceptable fit was obtained with 46% of Cu$_{0.2}$(Zn)InS$_2$/ZnS and 54% of CuIn$_{0.4}$S$_2$. This result reveals that, at most, Cu$_{0.3}$(Zn)InS$_2$ has half the amount of S on a ZnS shell than Cu$_{0.2}$(Zn)InS$_2$/ZnS. We also attribute the difficulty in obtaining high quality fit results using either linear combination fitting of the other QDs or a reconstruction from principal component analysis, to an imperfect shell formed in the case of Cu$_{0.3}$(Zn)InS$_2$ as compared to Cu$_{0.2}$(Zn)InS$_2$/ZnS, which corresponds to a fully passivated case. Nonetheless, this method stands as a more direct and reliable way to detect the presence of a surface shell layer and to study the thickness of the formed ZnS shell compared to TEM, where it is difficult to distinguish the layers, requiring a very high resolution and the combination with spectroscopy.[43] Furthermore, we can extract an approximate spectrum of S in the ZnS environment by carrying out different subtractions, obtaining a good agreement with the reference (Fig. S7). Using the results in Figs. S6 and S7 we can obtain rough estimates of the shell thickness (Table S3), obtaining values of 0.44 and 0.19 nm for Cu$_{0.2}$(Zn)InS$_2$/ZnS and Cu$_{0.3}$(Zn)InS$_2$. Comparing these values to the unit cell size of 0.54 nm we can deduce that Cu$_{0.2}$(Zn)InS$_2$/ZnS will have a complete passivation of the surface while for Cu$_{0.3}$(Zn)InS$_2$ it is only partial.

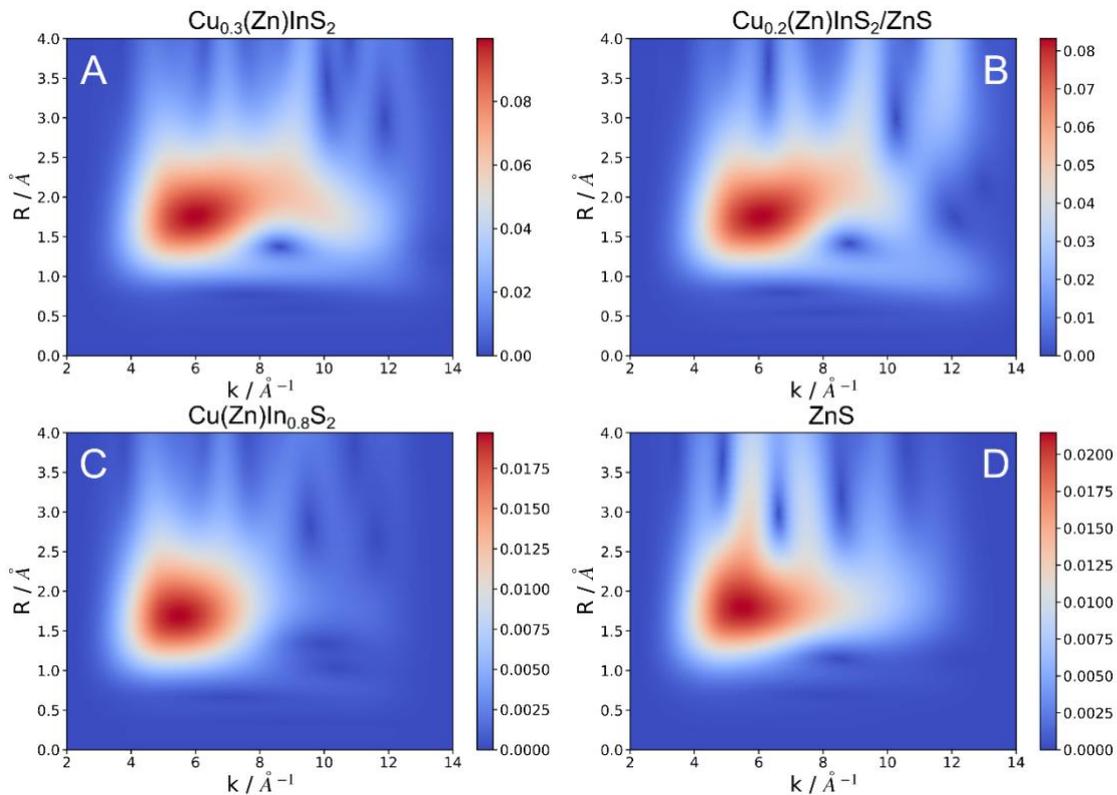

**Figure 3:** Wavelet transform of the Zn K-edge EXAFS of Cu$_{0.3}$(Zn)InS$_2$ (A), Cu$_{0.2}$(Zn)InS$_2$/ZnS (B), Cu(Zn)In$_{0.8}$S$_2$ (C), and ZnS (D).

So far, we have established that the $Zn^{2+}$ cations introduced during the synthesis form ZnS shells with different thicknesses for both $Cu_{0.3}(Zn)InS_2$ and $Cu_{0.2}(Zn)InS_2/ZnS$. However, we can obtain further details on their incorporation by performing more advanced analysis of the different Zn K-edge EXAFS signals using the wavelet transform (WT)[36] (Fig. 3). A WT is similar to the Fourier transform usually carried out in EXAFS (Fig. 2.B), but yields a 2D result resolved at which photoelectron wavevector ($k$) specific radial distance ($R$) appear. This is achieved using wavelets that serve as kernels for the integral transformation, effectively windowing it to a certain $k$ region. This transformation is then carried out varying the center $k$ of the wavelet in order to obtain the maps shown in Fig. 3.[44] This procedure can help to highlight EXAFS features and distinguish between photoelectron scattering with different elements, as heavier elements scatter at higher values of $k$.[36,44–46]

This gives us further insight into the incorporation of Zn into the core of the particles. Both $Cu_{0.3}(Zn)InS_2$ and $Cu_{0.2}(Zn)InS_2/ZnS$ exhibit the most complex spectra, which is reminiscent of a mixture of the map for $Cu(Zn)In_{0.8}S_2$ and that of ZnS. Once again, the similarity with ZnS shows the formation of a shell of this material in both QD samples. Furthermore, both Cu-deficient samples present a weak signal around R=1 Å and $k$ = 10-12 Å$^{-1}$. When analyzing wavelet transforms of EXAFS data, the value of $k$ is related to the atomic mass of the scatterer.[45,46] Thus, while most of the signal can be related to the scattering with the light S atoms ($k$ = 5-7 Å$^{-1}$), this one would come from scattering off much heavier atoms, such as Zn, Cu, or In, at shorter distances. Therefore, we propose it originates from Zn atoms incorporated as interstitial defects, in closer proximity to those other metal atoms. Note that the effective ionic radius of $Zn^{2+}$ is 0.6 Å,[47] making interstitial distances of 1 Å plausible. Interestingly, $Cu(Zn)In_{0.8}S_2$, shows an even weaker signal from these defects. In this case, Zn must be mainly incorporated as a substituent of Cu or In. Indeed, previous studies considered that Zn was being introduced as a substituent, eventually forming a ZnS shell, after sufficient substitution of $Cu^+$ and $In^{3+}$ ions.[18] An alternative data treatment enhances these features in Fig. S8. It is worth noting that the same effect is observed for the wavelet transforms of the Cu K-edge EXAFS data (Fig. S9). Once again, the two samples that show an elongated signal around R=1 Å and $k$ = 10-12 Å$^{-1}$ are $Cu_{0.3}(Zn)InS_2$ and $Cu_{0.2}(Zn)InS_2/ZnS$. Therefore, when looking at the photoelectrons emitted from Cu atoms, we can also observe the effect of the closely lying interstitial Zn defects. Thus, they are also introduced in the core of the QDs for these two samples and are not only present at the ZnS shell. Furthermore, these two samples, and to a lesser extent $Cu_{0.3}InS_2$, show a clear scattering at k = 10-12 Å$^{-1}$ over larger distances, indicating a higher presence of other heavy atoms (Cu, In, Zn) at the typical Cu-S distance. The accumulation at the highest $k$ values ($k$ = 12 Å$^{-1}$) compared to the weaker signal at R = 1 Å, and the absence at the Zn K-edge, may suggest that this signal stems from interstitial $In^{3+}$ ions that approach the Cu position in Cu-deficient samples. This analysis underlines the importance of measuring multi-edge XAS data, especially for composite materials such as QDs, which facilitates more advanced structural analysis by correlating signals originating from different atomic constituents of the particle, both in the core and on the surface of it.

**Implications of structural modifications on charge carrier dynamics and surface passivation**

In order to further assess the level and nature of surface passivation in the two different Zn-doped Cu-deficient samples used in our studies, we have complemented our static PL measurements with transient absorption spectroscopy (TAS) and time-resolved photoluminescence (TRPL) studies using additional electron (benzoquinone, BQ), and hole (phenothiazine, PTZ) acceptors. The former technique allowed us to observe the overall quenching of the PL signal, while the latter two allowed us to observe the decay kinetics of the photoexcited QDs on two different timescales: i) on ultrafast (<1ns) with TAS and ii) on nanosecond (>1ns) with TRPL. We carried out the PL measurements using the same PLQY spectrometer that was used to obtain the results shown in Fig. 1 and thus to maximize the accuracy. In steady-state PL, we observe a complete quenching of the signals with the addition of BQ, and a moderate effect with PTZ, mainly on $Cu_{0.3}(Zn)InS_2$ (top panels in Fig. 4 and Table 1). In time-resolved data, we observe a fast and efficient, ~ 97%, PL quenching by electron transfer using BQ for the $Cu_{0.2}(Zn)InS_2/ZnS$ sample. We can discard energy transfer because the absorption of BQ is of higher energy and does not overlap with the emission of CZIS QDs. The addition of BQ to the colloidal dispersion of $Cu_{0.3}(Zn)InS_2$ produced a chemical degradation and agglomeration, preventing us from carrying out any electron transfer study in it. Thus, the shell in this sample is insufficient to prevent chemical attacks beyond electron transfer. The effect of PTZ in the TRPL measurements is small in all cases, producing a quenching of the integrated TRPL of 4% and 13%, for $Cu_{0.2}(Zn)InS_2/ZnS$ and $Cu_{0.3}(Zn)InS_2$ respectively (Table 1). Slightly higher quenching efficiencies are obtained from TRPL lifetime fitting as seen in the SI (Section 7). Alternatively, since TAS is sensitive to the sum of the distribution function of electron and holes, and the ratio of effective masses is $m_h/m_e = 8$, TAS is not very sensitive to the dynamics of holes.[48,49] This explains why we observe a much larger effect with BQ and $Cu_{0.2}(Zn)InS_2/ZnS$, but only a small one is seen for $Cu_{0.3}(Zn)InS_2$ with PTZ.

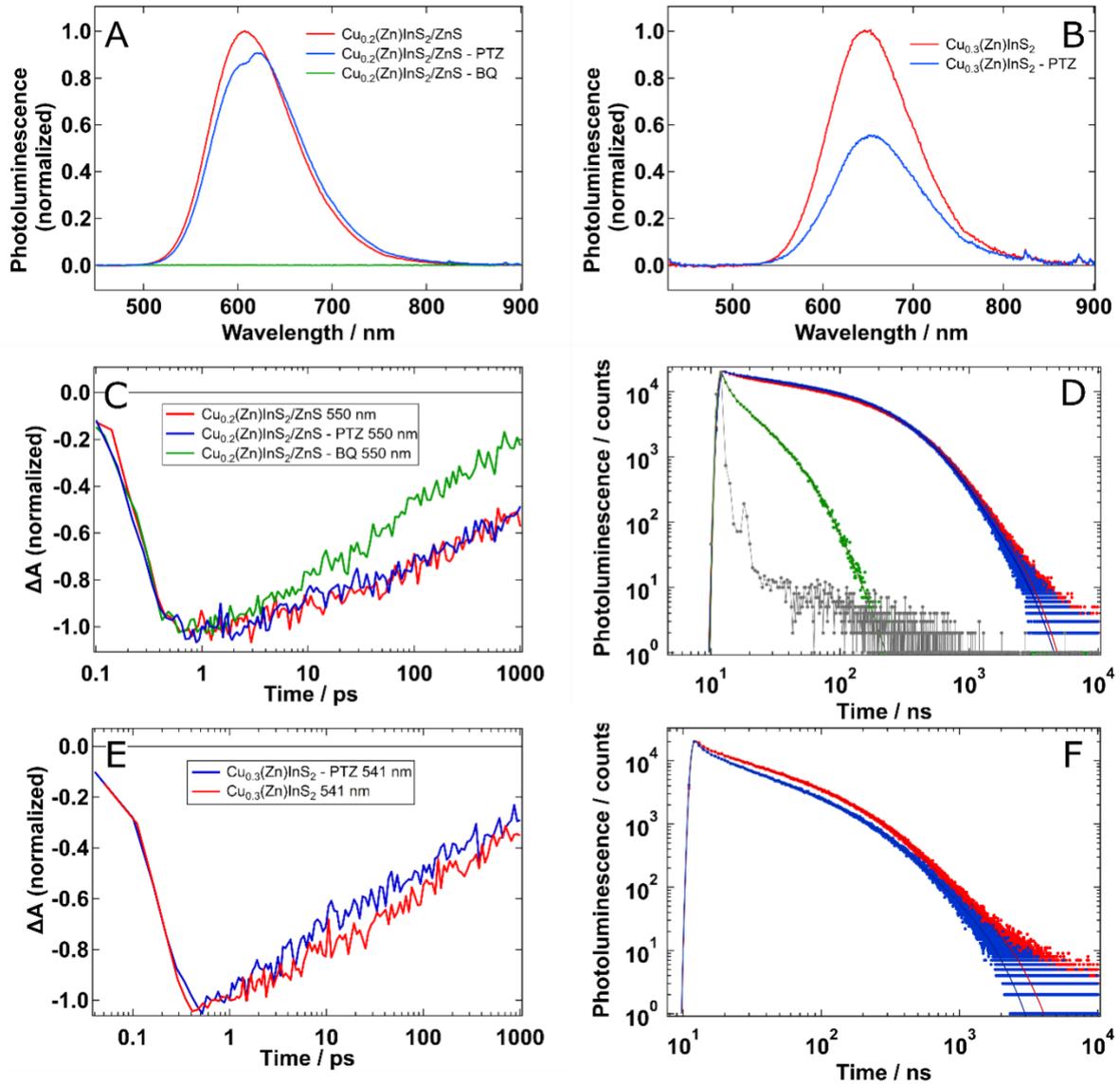

**Figure 4:** Top two panels: fluorescence quenching in $Cu_{0.2}(Zn)InS_2/ZnS$ (A) and $Cu_{0.3}(Zn)InS_2$ (B). Bottom four panels: transient absorption (left, $\lambda_{pump}$ = 520 nm, Fluence = 80 µJ·cm$^{-2}$) and time-resolved photoluminescence (right, $\lambda_{pump,}$ = 355 nm) measurements for $Cu_{0.2}(Zn)InS_2/ZnS$ (B and C) and $Cu_{0.3}(Zn)InS_2$ (E and F) with an electron acceptor (BQ, only for $Cu_{0.2}(Zn)InS_2/ZnS$) and a hole acceptor (PTZ). The instrument response function (IRF) is shown in grey. While the effect of the electron acceptor is clear, the effect of the hole acceptor seems limited on the time-resolved measurements, slightly observed in $Cu_{0.3}(Zn)InS_2$.

Taking into account the band alignment of $CuInS_2$ and ZnS, shown in Scheme 1.A, in addition to all the presented results, we can consider that: a) a shell of ZnS is being formed, which efficiently blocks hole and electron trapping at the surface in $Cu_{0.2}(Zn)InS_2/ZnS$, and to a slightly lower degree in $Cu_{0.3}(Zn)InS_2$; and b) this does not prevent all charge transfer to suitable acceptors. Indeed, because we observe an efficient electron transfer in $Cu_{0.2}(Zn)InS_2/ZnS$, electrons can, in principle, tunnel the ZnS barrier efficiently. Meanwhile, holes can also transfer in $Cu_{0.3}(Zn)InS_2$ due to its small shell thickness, whereas this effect is marginal in $Cu_{0.2}(Zn)InS_2/ZnS$. However, to understand the dramatic difference between electron and hole transfer dynamics, we need to consider the photophysical mechanisms in CIS QDs.

According to the most accepted model, after excitation at the band edge, an exciton is created. Then, the hole is trapped into a confined hole state (CHS) related to a Cu defect, and the emission occurs from the electron delocalized in the conduction band and the hole in the CHS.[50,51]

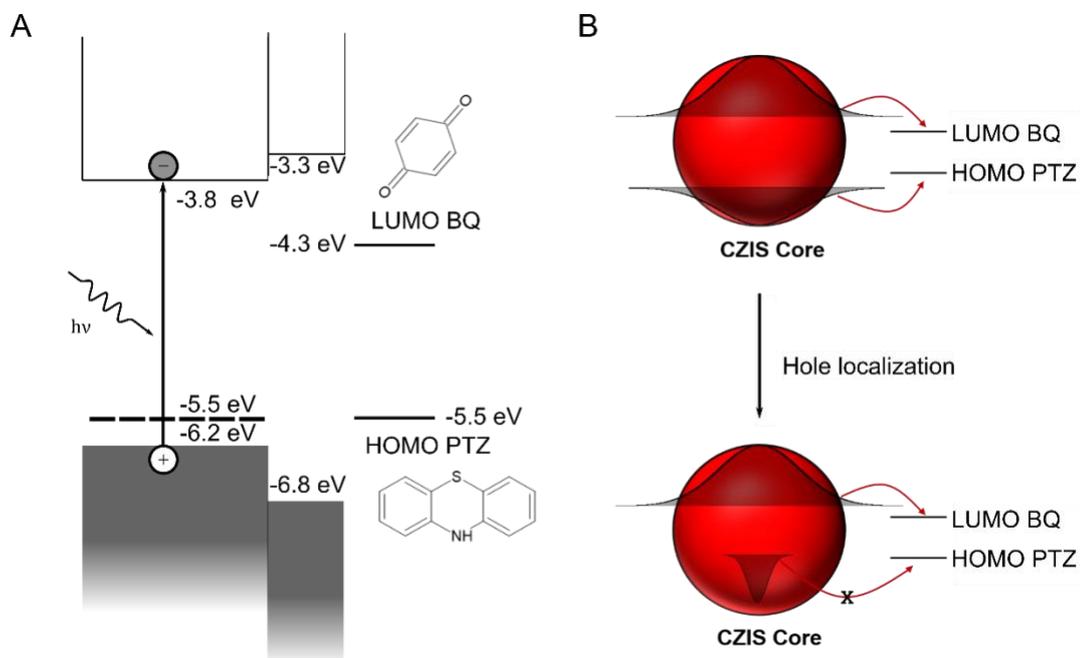

**Scheme 1:** A) Approximate band alignment between a 2.4 eV bandgap $CuInS_2$ core,[50] the ZnS shell,[52] and the molecular acceptors[53], according to the literature. The hole localization defects (CHS) lie at an energy of about -5.5 eV according to the same cyclic voltammetry study.[50] B) Representation of the difference in hole transfer between delocalized and localized holes.

Therefore, the electrons in the QD can be easily transferred to the acceptor because of a high overlap integral involving the QD CB and BQ LUMO wavefunctions. On the other hand, due to a lower overlap integral of the CHS orbitals with the PTZ HOMO, hole transfer in CIS is efficient only before the hole localization process. In other words, while the hole is still delocalized in the VB. This is illustrated in Scheme 1.B.

Furthermore, the energy levels of the CHS and the PTZ HOMO may be very similar, further limiting the driving force for charge transfer (Scheme 1.A). Finally, if the ZnS shell is thick enough, it will efficiently limit hole transfer, as seen for $Cu_{0.2}(Zn)InS_2/ZnS$, since it will act as a barrier for positive charge carriers.

The limitation of our data presented in Fig. 4 is that TAS is not very sensitive to holes in this case, and our TRPL measurements, which are equally sensitive to both charge carriers, have a time resolution of a few nanoseconds. Nonetheless, a change in the TAS decay for $Cu_{0.3}(Zn)InS_2$ with PTZ is observed. In addition, the TRPL charge transfer efficiency differs from that of the PL data for $Cu_{0.3}(Zn)InS_2$ (Table 1). Indeed, as shown earlier, PL measurements yield a hole transfer of about 40% for $Cu_{0.3}(Zn)InS_2$ and 3% for $Cu_{0.2}(Zn)InS_2/ZnS$. This discrepancy can be explained assuming that most of the observed TRPL signal originates from the population of $Cu_{0.3}(Zn)InS_2$ QDs that had undergone hole localization before hole transfer to PTZ could have occurred.

Meanwhile, the quenching of the PL signal corresponds to all the transferred holes. This is further evidenced by the wavelength-dependence of the quenching efficiency ($\Phi_q$) of the PL signals. As shown in Fig. S10, the higher the excess energy, the higher the $\Phi_q$. Therefore, photogenerated holes with higher temperatures will have higher probability of transferring to PTZ before cooling down and localizing into the CHS. In line with the above, the emission spectrum with PTZ shows a small red-shift, which can be explained by the effect of CHS states closer to the VB being more susceptible to hole transfer to PTZ, as it would be energetically more favorable (Fig. S11).

**Table 1:** Quenching by the electron and hole acceptors BQ and PTZ. While the effect of BQ is large, extracting many electrons, the effect of PTZ is more limited for $Cu_{0.2}(Zn)InS_2/ZnS$ and further decreases in TRPL measurements due to hole localization. The PL results are the average of the three measured excitation wavelengths (420, 490 and 510 nm). The TRPL $\Phi_q$ is calculated from the integration of the TRPL traces. It considers equal emission at *ca.* 1 ns and thus evaluates the quenching after that. Alternatively, $\Phi_q$ calculated from lifetime fitting is shown in Table S4.

| QD | QD-PTZ | QD-BQ |
|---|---|---|
| PL | $\Phi_q$ | $\Phi_q$ |
| $Cu_{0.2}(Zn)InS_2/ZnS$ | 0.06 | 1 |
| $Cu_{0.3}(Zn)InS_2$ | 0.4 | |
| TRPL | $\Phi_q$ | $\Phi_q$ |
| $Cu_{0.2}(Zn)InS_2/ZnS$ | 0.04 | 0.97 |
| $Cu_{0.3}(Zn)InS_2$ | 0.13 | |

In conclusion, the localized holes have a lower charge transfer efficiency, which is obtained from the TRPL lifetime quenching, and only the thin ZnS shell in $Cu_{0.3}(Zn)InS_2$ permits hole transfer before CHS formation. The small quenching in $Cu_{0.2}(Zn)InS_2/ZnS$ corresponds to a very slow, low probability, hole transfer captured by TRPL and static PL in a similar way with no ultrafast component. This is because the shell is too thick for the hole wavefunction to leak to the surface before CHS formation. Furthermore, according to our previous study [24] the samples without any Zn doping, $Cu_{0.3}InS_2$ and $CuIn_{0.4}S_2$, have a hole-trapping process unrelated to Cu, which is responsible for their significantly lower PLQY. A similar process would be occurring in $Cu(Zn)In_{0.8}S_2$, due to the lack of detectable PL signal and the absence of a ZnS shell.

Consequently, it is impossible to observe the PTZ quenching in $Cu(Zn)In_{0.8}S_2$ due to the competition with ligand or surface state hole and electron trapping processes, affecting a large proportion of the photogenerated charge carriers.

The values for band edges and acceptors are taken from literature data of similar QDs,[50,53] and are subject to changes due to the exact size of the core and the shell.[52–55] For example, smaller bandgap QDs may increase the VB limit (up to *ca.* -5.1 eV for bulk). Alternatively, electrons may easily tunnel through the few nanometer-thick shell because of their smaller effective mass.[56] However, our observations allow us to conclude that the synthesis of $Cu_{0.2}(Zn)InS_2/ZnS$ produces a sufficiently thick ZnS shell to i) efficiently passivate against hole trapping, ii) extend the lifetime for recombination, while iii) still allow the flow of electrons for charge transfer processes outside the QD, while only a few holes can be transferred over long timescales. Alternatively, the $Cu_{0.3}(Zn)InS_2$ synthesis, without thermal post-treatment, provides a sufficient passivation to allow for much more efficient hole transfer, both fast and slow (free and localized holes). Another explanation for the reduced hole transfer compared to the electrons is the formation of a type II band alignment.[54] Furthermore, we learned from these results that electron trapping is not a limiting process in the recombination of Cu-deficient CZIS, as electrons are not efficiently passivated from transfer toward the surface. The same cannot be said about the Cu-rich samples, since we have observed a considerable increase in the lattice disorder as corroborated by the Cu K-edge XAS spectra, which would translate into a larger number of internal trap states.[24]

**Conclusions**

We have carried out a systematic study of CIS QDs with different levels of Zn doping and stoichiometry combining optical and X-ray spectroscopies. We have observed that Zn can be incorporated both inside and around the core of the nanoparticle, forming a shell of ZnS. In Cu-deficient samples, $Zn^{2+}$ ions form a ZnS shell when added during the synthesis without the application of any post-treatment of approximately half the volume compared to the shell growth treatment. The S K-edge is a powerful probe for in-situ monitoring shell growth in colloidal dispersions of QDs. Furthermore, Zn K-edge measurements present signals compatible with interstitial defect in Cu-deficient samples, although it does not have detrimental effects on the photophysics of the QDs.

In Cu-rich samples, $Zn^{2+}$ is incorporated mainly as a substituent of $Cu^+$ or $In^{3+}$. We have shown the strength of XAS at the Zn and S K-edges to determine Zn incorporation and ZnS formation, which is not easily observable with TEM due to the similarity with the core. Furthermore, we presented how wavelet transforms of EXAFS data can help identifying interstitial defects.

From quenching studies with electron and hole scavengers, we have determined the formation of a ZnS shell, which prevents hole trapping at the surface, but it can be tailored to allow its transfer to an acceptor. On the other hand, electrons can easily be transferred through the barrier to an appropriate acceptor. The difference in transfer efficiency can be related to the localization of holes in the core and is, thus, compatible with the existence of CHS close to the VB edge.

**Acknowledgements**

The steady-state XAS experiments were performed at the CLÆSS beamline at ALBA Synchrotron within the Proposal 2021095311. We would like to greatly acknowledge the help and support for CLÆSS beamline staff in preparation and execution of XAS experiments, which contributed to the results presented here. The authors would also like to thank Dr. Reinhold Wannemacher and Luis Colmenar for their help with TRPL and PLQY measurements, respectively, at IMDEA Nanoscience.

AB is grateful to the Spanish "Ministerio de Universidades" and the "Plan de Recuperación, Transformación y Resiliencia", as well as the UAM, for his "Margarita Salas" grant (ref. CA1/RSUE/2021-00809).

In addition, he receives funding from the European Union's Horizon 2020 research and innovation programme under the Marie Sklodowska-Curie agreement No. 101034431 and from the "Severo Ochoa" Programme for Centres of Excellence in R&D (CEX2020-001039S / AEI / 10.13039/501100011033. WG acknowledges funding from Spanish Ministry of Universities through "Ayudas Beatriz Galindo" (BEAGAL18/00092), Regional Government of Madrid and Universidad Autónoma de Madrid through "Proyectos de I+D para Investigadores del Programa Beatriz Galindo" grant (Ref. SI2/PBG/2020-00003) and from Spanish Ministry of Science, Innovation and Universities through "Proyectos de I+D+i 2019" grant (Ref. PID2019-108678GB-I00) and "Proyectos de I+D+i 2022" grant (Ref. PID2022-140257NB-I00). BRCV and LAP thank São Paulo Research Foundation, FAPESP, under grants 2018/15574-6 and 2022/06470-8. BRCV also thanks FAPESP for the postdoctoral scholarship under grants 2020/16077-6 and 2024/01722-4. AFVF thanks FAPESP under the grant 2023/10395-4. AFN acknowledges the support from the FAPESP (grant no. 2017/11986-5) and Shell and the strategic importance of the support given by ANP (Brazil's National Oil, Natural Gas, and Biofuels Agency). MAS acknowledges the support from the CNPq and FAPEMIG (grant no. APQ-02598-23). JC-G acknowledges the MICINN-FEDER (No. PID2021-128313OB-I00), support from the Regional Government of Madrid (NMAT2D-CM), a Research Consolidation Grant (No. CNS2022-36191) and project PDC202-314587-1I00 from the Spanish Ministry of Science and Innovation. VVM acknowledges grants TED2021-131906A-100 and RYC2022-035200-I funded by Spanish Ministry of Science, Innovation and Universities (10.13039/501100011033) and support from the Regional Government of Madrid (2019-T2/IND-12737 and 2024-T1/TEC-31349). SGO is grateful to the Spanish Ministry of Science and Innovation for a Ph.D. grant (FPI, PRE2019-09345). This work was also partially funded by the regional government of Madrid (Spain) through the Tecnologias 2024 program, project MATRIX-CM (TEC-2024/TEC-85).

# References

(1) Long, Z.; Zhang, W.; Tian, J.; Chen, G.; Liu, Y.; Liu, R. Recent Research on the Luminous Mechanism, Synthetic Strategies, and Applications of CuInS2 Quantum Dots. *Inorg. Chem. Front.* **2021**, *8* (4), 880–897. https://doi.org/10.1039/D0QI01228A.

(2) Wang, Z.; Zhang, X.; Xin, W.; Yao, D.; Liu, Y.; Zhang, L.; Liu, W.; Zhang, W.; Zheng, W.; Yang, B.; Zhang, H. Facile Synthesis of Cu–In–S/ZnS Core/Shell Quantum Dots in 1-Dodecanethiol for Efficient Light-Emitting Diodes with an External Quantum Efficiency of 7.8%. *Chem. Mater.* **2018**, *30* (24), 8939–8947. https://doi.org/10.1021/acs.chemmater.8b04282.

(3) Kim, N.; Na, W.; Yin, W.; Jin, H.; Ahn, T. K.; Cho, S. M.; Chae, H. CuInS2/ZnS Quantum Dot-Embedded Polymer Nanofibers for Color Conversion Films. *J. Mater. Chem. C* **2016**, *4* (13), 2457–2462. https://doi.org/10.1039/C5TC03967C.

(4) Bergren, M. R.; Makarov, N. S.; Ramasamy, K.; Jackson, A.; Guglielmetti, R.; McDaniel, H. High-Performance CuInS$_2$ Quantum Dot Laminated Glass Luminescent Solar Concentrators for Windows. *ACS Energy Lett.* **2018**, *3* (3), 520–525. https://doi.org/10.1021/acsenergylett.7b01346.

(5) Anand, A.; Zaffalon, M. L.; Gariano, G.; Camellini, A.; Gandini, M.; Brescia, R.; Capitani, C.; Bruni, F.; Pinchetti, V.; Zavelani-Rossi, M.; Meinardi, F.; Crooker, S. A.; Brovelli, S. Evidence for the Band-Edge Exciton of CuInS$_2$ Nanocrystals Enables Record Efficient Large-Area Luminescent Solar Concentrators. *Adv. Funct. Mater.* **2020**, *30* (4), 1906629. https://doi.org/10.1002/adfm.201906629.

(6) Gungor, K.; Du, J.; Klimov, V. I. General Trends in the Performance of Quantum Dot Luminescent Solar Concentrators (LSCs) Revealed Using the "Effective LSC Quality Factor." *ACS Energy Lett.* **2022**, *7* (5), 1741–1749. https://doi.org/10.1021/acsenergylett.2c00781.

(7) Chiang, Y.-H.; Lin, K.-Y.; Chen, Y.-H.; Waki, K.; Abate, M. A.; Jiang, J.-C.; Chang, J.-Y. Aqueous Solution-Processed off-Stoichiometric Cu–In–S QDs and Their Application in Quantum Dot-Sensitized Solar Cells. *J. Mater. Chem. A* **2018**, *6* (20), 9629–9641. https://doi.org/10.1039/C8TA01064A.

(8) Du, J.; Du, Z.; Hu, J.-S.; Pan, Z.; Shen, Q.; Sun, J.; Long, D.; Dong, H.; Sun, L.; Zhong, X.; Wan, L.-J. Zn–Cu–In–Se Quantum Dot Solar Cells with a Certified Power Conversion Efficiency of 11.6%. *J. Am. Chem. Soc.* **2016**, *138* (12), 4201–4209. https://doi.org/10.1021/jacs.6b00615.

(9) Gao, X.; Liu, X.; Lin, Z.; Liu, S.; Su, X. CuInS2 Quantum Dots as a Near-Infrared Fluorescent Probe for Detecting Thrombin in Human Serum. *Analyst* **2012**, *137* (23), 5620–5624. https://doi.org/10.1039/C2AN35888C.

(10) Liu, S.; Shi, F.; Chen, L.; Su, X. Dopamine Functionalized CuInS2 Quantum Dots as a Fluorescence Probe for Urea. *Sens. Actuators B Chem.* **2014**, *191*, 246–251. https://doi.org/10.1016/j.snb.2013.09.056.

(11) Deng, D.; Chen, Y.; Cao, J.; Tian, J.; Qian, Z.; Achilefu, S.; Gu, Y. High-Quality CuInS2/ZnS Quantum Dots for In Vitro and In Vivo Bioimaging. *Chem. Mater.* **2012**, *24* (15), 3029–3037. https://doi.org/10.1021/cm3015594.

(12) Foda, M. F.; Huang, L.; Shao, F.; Han, H.-Y. Biocompatible and Highly Luminescent Near-Infrared CuInS2/ZnS Quantum Dots Embedded Silica Beads for Cancer Cell Imaging. *ACS Appl. Mater. Interfaces* **2014**, *6* (3), 2011–2017. https://doi.org/10.1021/am4050772.

(13) Grabolle, M.; Spieles, M.; Lesnyak, V.; Gaponik, N.; Eychmüller, A.; Resch-Genger, U. Determination of the Fluorescence Quantum Yield of Quantum Dots: Suitable Procedures and Achievable Uncertainties. *Anal. Chem.* **2009**, *81* (15), 6285–6294. https://doi.org/10.1021/ac900308v.

(14) Shockley, W.; Read Jr, W. T. Statistics of the Recombinations of Holes and Electrons. *Phys. Rev.* **1952**, *87* (5), 835.

(15) Vasudevan, D.; Gaddam, R. R.; Trinchi, A.; Cole, I. Core–Shell Quantum Dots: Properties and Applications. *J. Alloys Compd.* **2015**, *636*, 395–404. https://doi.org/10.1016/j.jallcom.2015.02.102.

(16) Vega-Mayoral, V.; Garcia-Orrit, S.; Wang, P.; Morales-Márquez, R.; Rodríguez, E. M.; Juárez, B. H.; Cabanillas-Gonzalez, J. Exploring Many-Body Phenomena: Biexciton Generation and Auger Recombination in Ag2S-Based Nanocrystals. *Nanoscale* **2025**. https://doi.org/10.1039/D5NR00511F.

(17) Wu, K.; Liang+, G.; Kong, D.; Chen, J.; Chen, Z.; Shan, X.; McBride, J. R.; Lian, T. Quasi-Type II CuInS2/CdS Core/Shell Quantum Dots. *Chem. Sci.* **2016**, *7* (2), 1238–1244. https://doi.org/10.1039/C5SC03715H.

(18) Park, J.; Kim, S.-W. CuInS2/ZnS Core/Shell Quantum Dots by Cation Exchange and Their Blue-Shifted Photoluminescence. *J. Mater. Chem.* **2011**, *21* (11), 3745–3750. https://doi.org/10.1039/C0JM03194A.

(19) Berends, A. C.; van der Stam, W.; Hofmann, J. P.; Bladt, E.; Meeldijk, J. D.; Bals, S.; de Mello Donega, C. Interplay between Surface Chemistry, Precursor Reactivity, and Temperature Determines Outcome of ZnS Shelling Reactions on CuInS2 Nanocrystals. *Chem. Mater.* **2018**, *30* (7), 2400–2413. https://doi.org/10.1021/acs.chemmater.8b00477.

(20) Moser, A.; Yarema, M.; Lin, W. M. M.; Yarema, O.; Yazdani, N.; Wood, V. In Situ Monitoring of Cation-Exchange Reaction Shell Growth on Nanocrystals. *J. Phys. Chem. C* **2017**, *121* (43), 24345–24351. https://doi.org/10.1021/acs.jpcc.7b08571.

(21) Vale, B. R. C.; Socie, E.; Cunha, L. R. C.; Fonseca, A. F. V.; Vaz, R.; Bettini, J.; Moser, J.-E.; Schiavon, M. A. Revealing Exciton and Metal–Ligand Conduction Band Charge Transfer Absorption Spectra in Cu-Zn-In-S Nanocrystals. *J. Phys. Chem. C* **2020**, *124* (50), 27858–27866. https://doi.org/10.1021/acs.jpcc.0c09681.

(22) Aldakov, D.; Reiss, P. Safer-by-Design Fluorescent Nanocrystals: Metal Halide Perovskites vs Semiconductor Quantum Dots. *J. Phys. Chem. C* **2019**, *123* (20), 12527–12541. https://doi.org/10.1021/acs.jpcc.8b12228.

(23) Yano, J.; Yachandra, V. K. X-Ray Absorption Spectroscopy. *Photosynth. Res.* **2009**, *102* (2–3), 241–254. https://doi.org/10.1007/s11120-009-9473-8.

(24) Burgos-Caminal, A.; Vale, B. R. C.; Fonseca, A. F. V.; Collet, E. P. P.; Hidalgo, J. F.; García, L.; Watson, L.; Borrell-Grueiro, O.; Corrales, M. E.; Choi, T.-K.; Katayama, T.; Fan, D.; Vega-Mayoral, V.; García-Orrit, S.; Nozawa, S.; Penfold, T. J.; Cabanillas-Gonzalez, J.; Adachi, S.-I.; Bañares, L.; Nogueira, A. F.; Padilha, L. A.; Schiavon, M. A.; Gawelda, W. Selective Tracking of Charge Carrier Dynamics in CuInS2 Quantum Dots. arXiv December 19, 2024. https://doi.org/10.48550/arXiv.2412.15418.

(25) Sun, C.; Bai, L.; Roldao, J. C.; Burgos-Caminal, A.; Borrell-Grueiro, O.; Lin, J.; Huang, W.; Gierschner, J.; Gawelda, W.; Bañares, L.; Cabanillas-González, J. Boosting the Stimulated Emission Properties of Host:Guest Polymer Blends by Inserting Chain Twists in the Host Polymer. *Adv. Funct. Mater.* **2022**, *32* (48), 2206723. https://doi.org/10.1002/adfm.202206723.

(26) Simonelli, L.; Marini, C.; Olszewski, W.; Ávila Pérez, M.; Ramanan, N.; Guilera, G.; Cuartero, V.; Klementiev, K. CLÆSS: The Hard X-Ray Absorption Beamline of the ALBA CELLS Synchrotron. *Cogent Phys.* **2016**, *3* (1), 1231987. https://doi.org/10.1080/23311940.2016.1231987.

(27) Nelson, H. D.; Gamelin, D. R. Valence-Band Electronic Structures of $Cu^+$-Doped ZnS, Alloyed Cu–In–Zn–S, and Ternary $CuInS_2$ Nanocrystals: A Unified Description of Photoluminescence across Compositions. *J. Phys. Chem. C* **2018**, *122* (31), 18124–18133. https://doi.org/10.1021/acs.jpcc.8b05286.

(28) Momma, K.; Izumi, F. Vesta 3 for Three-Dimensional Visualization of Crystal, Volumetric and Morphology Data. *J. Appl. Crystallogr.* **2011**, *44* (6), 1272–1276. https://doi.org/10.1107/S0021889811038970.

(29) Hahn, H.; Frank, G.; Klingler, W.; Meyer, A.-D.; Störger, G. Untersuchungen über ternäre Chalkogenide. V. Über einige ternäre Chalkogenide mit Chalkopyritstruktur. *Z. Für Anorg. Allg. Chem.* **1953**, *271* (3–4), 153–170. https://doi.org/10.1002/zaac.19532710307.

(30) Jumpertz, E. A. Electron-Density Distribution in Zinc Blende. *Z. Für Elektrochem. Angew. Phys. Chem.* **1955**, No. 59, 419–425.

(31) Kurik, M. V. Urbach Rule. *Phys. Status Solidi A* **1971**, *8* (1), 9–45. https://doi.org/10.1002/pssa.2210080102.

(32) Sakamoto, M.; Chen, L.; Okano, M.; Tex, D. M.; Kanemitsu, Y.; Teranishi, T. Photoinduced Carrier Dynamics of Nearly Stoichiometric Oleylamine-Protected Copper Indium Sulfide Nanoparticles and Nanodisks. *J. Phys. Chem. C* **2015**, *119* (20), 11100–11105. https://doi.org/10.1021/jp511864p.

(33) Turnbull, M. J.; Vaccarello, D.; Wong, J.; Yiu, Y. M.; Sham, T.-K.; Ding, Z. Probing the CZTS/CdS Heterojunction Utilizing Photoelectrochemistry and x-Ray Absorption Spectroscopy. *J. Chem. Phys.* **2018**, *148* (13), 134702. https://doi.org/10.1063/1.5016351.

(34) Wende, H.; Baberschke, K. Atomic EXAFS: Evidence for Photoelectron Backscattering by Interstitial Charge Densities. *J. Electron Spectrosc. Relat. Phenom.* **1999**, *101–103*, 821–826. https://doi.org/10.1016/S0368-2048(98)00431-9.

(35) Wende, H.; Litwinski, C.; Scherz, A.; Gleitsmann, T.; Li, Z.; Sorg, C.; Baberschke, K.; Ankudinov, A.; Rehr, J. J.; Jung, C. A Systematic Study of Embedded Atom EXAFS: The (2 1)O/Cu(110) Reconstruction as an Ideal Prototype System. *J. Phys. Condens. Matter* **2003**, *15* (30), 5197–5206. https://doi.org/10.1088/0953-8984/15/30/302.

(36) Penfold, T. J.; Tavernelli, I.; Milne, C. J.; Reinhard, M.; Nahhas, A. E.; Abela, R.; Rothlisberger, U.; Chergui, M. A Wavelet Analysis for the X-Ray Absorption Spectra of Molecules. *J. Chem. Phys.* **2013**, *138* (1), 014104. https://doi.org/10.1063/1.4772766.

(37) Ravel, B.; Newville, M. ATHENA, ARTEMIS, HEPHAESTUS: Data Analysis for X-Ray Absorption Spectroscopy Using IFEFFIT. *J. Synchrotron Radiat.* **2005**, *12* (4), 537–541. https://doi.org/10.1107/S0909049505012719.

(38) McCubbin Stepanic, O.; Ward, J.; Penner-Hahn, J. E.; Deb, A.; Bergmann, U.; DeBeer, S. Probing a Silent Metal: A Combined X-Ray Absorption and Emission Spectroscopic Study of Biologically Relevant Zinc Complexes. *Inorg. Chem.* **2020**, *59* (18), 13551–13560. https://doi.org/10.1021/acs.inorgchem.0c01931.

(39) Tapley, A.; Liu, L.; Cui, X.; Zuin, L.; Love, D. A.; Zhou, J.; Sham, T.-K.; Ding, Z. Assessing the Band Structure of CuInS2 Nanocrystals and Their Bonding with the Capping Ligand. *J. Phys. Chem. C* **2015**, *119* (36), 20967–20974. https://doi.org/10.1021/acs.jpcc.5b05940.

(40) Ye, K.; Siah, S. C.; Erslev, P. T.; Akey, A.; Settens, C.; Hoque, M. S. B.; Braun, J.; Hopkins, P.; Teeter, G.; Buonassisi, T.; Jaramillo, R. Tuning Electrical, Optical, and Thermal Properties through Cation Disorder in Cu2ZnSnS4. *Chem. Mater.* **2019**, *31* (20), 8402–8412. https://doi.org/10.1021/acs.chemmater.9b02287.

(41) Yamazoe, S.; Kou, H.; Wada, T. A Structural Study of Cu–In–Se Compounds by x-Ray Absorption Fine Structure. *J. Mater. Res.* **2011**, *26* (12), 1504–1516. https://doi.org/10.1557/jmr.2011.63.

(42) Cho, D.-Y.; Xi, L.; Boothroyd, C.; Kardynal, B.; Lam, Y. M. The Role of Ion Exchange in the Passivation of In(Zn)P Nanocrystals with ZnS. *Sci. Rep.* **2016**, *6* (1), 22818. https://doi.org/10.1038/srep22818.

(43) Weigert, F.; Müller, A.; Häusler, I.; Geißler, D.; Skroblin, D.; Krumrey, M.; Unger, W.; Radnik, J.; Resch-Genger, U. Combining HR-TEM and XPS to Elucidate the Core–Shell Structure of Ultrabright CdSe/CdS Semiconductor Quantum Dots. *Sci. Rep.* **2020**, *10* (1), 20712. https://doi.org/10.1038/s41598-020-77530-z.

(44) Funke, H.; Scheinost, A. C.; Chukalina, M. Wavelet Analysis of Extended X-Ray Absorption Fine Structure Data. *Phys. Rev. B* **2005**, *71* (9), 094110. https://doi.org/10.1103/PhysRevB.71.094110.

(45) Timoshenko, J.; Kuzmin, A. Wavelet Data Analysis of EXAFS Spectra. *Comput. Phys. Commun.* **2009**, *180* (6), 920–925. https://doi.org/10.1016/j.cpc.2008.12.020.

(46) Xia, Z.; Zhang, H.; Shen, K.; Qu, Y.; Jiang, Z. Wavelet Analysis of Extended X-Ray Absorption Fine Structure Data: Theory, Application. *Phys. B Condens. Matter* **2018**, *542*, 12–19. https://doi.org/10.1016/j.physb.2018.04.039.

(47) Shannon, R. D. Revised Effective Ionic Radii and Systematic Studies of Interatomic Distances in Halides and Chalcogenides. *Acta Crystallogr. A* **1976**, *32* (5), 751–767. https://doi.org/10.1107/S0567739476001551.

(48) Berends, A. C.; Mangnus, M. J. J.; Xia, C.; Rabouw, F. T.; de Mello Donega, C. Optoelectronic Properties of Ternary I–III–VI2 Semiconductor Nanocrystals: Bright Prospects with Elusive

Origins. *J. Phys. Chem. Lett.* **2019**, *10* (7), 1600–1616. https://doi.org/10.1021/acs.jpclett.8b03653.

(49) Morgan, D. P.; Kelley, D. F. What Does the Transient Absorption Spectrum of CdSe Quantum Dots Measure? *J. Phys. Chem. C* **2020**, *124* (15), 8448–8455. https://doi.org/10.1021/acs.jpcc.0c02566.

(50) Fuhr, A. S.; Yun, H. J.; Makarov, N. S.; Li, H.; McDaniel, H.; Klimov, V. I. Light Emission Mechanisms in $CuInS_2$ Quantum Dots Evaluated by Spectral Electrochemistry. *ACS Photonics* **2017**, *4* (10), 2425–2435. https://doi.org/10.1021/acsphotonics.7b00560.

(51) van der Stam, W.; de Graaf, M.; Gudjonsdottir, S.; Geuchies, J. J.; Dijkema, J. J.; Kirkwood, N.; Evers, W. H.; Longo, A.; Houtepen, A. J. Tuning and Probing the Distribution of $Cu^+$ and $Cu^{2+}$ Trap States Responsible for Broad-Band Photoluminescence in $CuInS_2$ Nanocrystals. *ACS Nano* **2018**, *12* (11), 11244–11253. https://doi.org/10.1021/acsnano.8b05843.

(52) Cerdán-Pasarán, A.; Esparza, D.; Zarazúa, I.; Reséndiz, M.; López-Luke, T.; De la Rosa, E.; Fuentes-Ramírez, R.; Alatorre-Ordaz, A.; Martínez-Benítez, A. Photovoltaic Study of Quantum Dot-Sensitized TiO2/CdS/ZnS Solar Cell with P3HT or P3OT Added. *J. Appl. Electrochem.* **2016**, *46* (9), 975–985. https://doi.org/10.1007/s10800-016-0972-y.

(53) Wu, K.; Liang, G.; Shang, Q.; Ren, Y.; Kong, D.; Lian, T. Ultrafast Interfacial Electron and Hole Transfer from CsPbBr3 Perovskite Quantum Dots. *J. Am. Chem. Soc.* **2015**, *137* (40), 12792–12795. https://doi.org/10.1021/jacs.5b08520.

(54) Liu, L.; Li, H.; Liu, Z.; Xie, Y.-H. The Conversion of CuInS2/ZnS Core/Shell Structure from Type I to Quasi-Type II and the Shell Thickness-Dependent Solar Cell Performance. *J. Colloid Interface Sci.* **2019**, *546*, 276–284. https://doi.org/10.1016/j.jcis.2019.03.075.

(55) Lv, M.; Zhu, J.; Huang, Y.; Li, Y.; Shao, Z.; Xu, Y.; Dai, S. Colloidal CuInS2 Quantum Dots as Inorganic Hole-Transporting Material in Perovskite Solar Cells. *ACS Appl. Mater. Interfaces* **2015**, *7* (31), 17482–17488. https://doi.org/10.1021/acsami.5b05104.

(56) Sylvia, S. S.; Park, H.-H.; Khayer, M. A.; Alam, K.; Klimeck, G.; Lake, R. K. Material Selection for Minimizing Direct Tunneling in Nanowire Transistors. *IEEE Trans. Electron Devices* **2012**, *59* (8), 2064–2069. https://doi.org/10.1109/TED.2012.2200688.

# Supplementary Information

# Unveiling Zn incorporation in CuInS$_2$ quantum dots: X-ray and optical analysis of doping effects, structural modifications and surface passivation

Andrés Burgos-Caminal[1,2*], Brener R. C. Vale[3,4], André F. V. Fonseca[6,4], Juan F. Hidalgo[1], Elisa P. P. Collet[1], Lázaro García,[2] Víctor Vega-Mayoral[1], Saül Garcia-Orrit[1], Iciar Arnay[1], Juan Cabanillas-González[1], Laura Simonelli[5], Ana Flávia Nogueira[6], Marco Antônio Schiavon[4], Thomas J. Penfold[7], Lazaro A. Padilha[3] and Wojciech Gawelda[2,1,8*]

*1. Madrid Institute for Advanced Studies IMDEA Nanoscience, Ciudad Universitaria de Cantoblanco, Calle Faraday 9, 28049 Madrid, Spain*
*2. Departamento de Química, Universidad Autónoma de Madrid, Ciudad Universitaria de Cantoblanco, Calle Francisco Tomás y Valiente 7, 28049 Madrid, Spain*
*3. Ultrafast laboratory spectroscopy, "Gleb Wataghin" Institute of Physics, University of Campinas, Brazil*
*4. Grupo de Pesquisa Química de Materiais, Departamento de Ciências Naturais, Universidade Federal de São João Del-Rei, Brazil*
*5. CELLS-ALBA Synchrotron Light Source, 08290 Cerdanyola del Vallès, Barcelona, Spain*
*6. Laboratório de Nanotecnologia e Energia Solar, Chemistry Institute, University of Campinas – UNICAMP, Campinas, São Paulo, Brazil*
*7. Chemistry, School of Natural and Environmental Sciences Newcastle University, NE1 7RU Newcastle upon Tyne, UK*
*8. Faculty of Physics, Adam Mickiewicz University, ul. Uniwersytetu Poznańskiego 2, 61-614 Poznań, Poland*

*Corresponding author(s): andres.burgos@imdea.org ; wojciech.gawelda@uam.es*

# 1. Quantum dot characterization

We characterized the physio-chemical properties of as-synthesized samples with three different techniques:

A) X-ray diffraction (XRD) of the samples deposited on a glass slide. We used a commercial Rigaku SmartLab SE multipurpose X-ray diffractometer. We employed a Bragg-Bentano geometry with Cu Kα (1.54 Å) as the source.

B) High-resolution transmission electron microscopy (HR-TEM) using a JEM-2100 from JEOL (Fig. S1).

C) Elemental analysis through X-ray fluorescence (XRF). We used an S2 PICOFOX from Bruker, obtaining the relative mass for each element.

**Table S1**. Elemental analysis of the stoichiometry through X-ray fluorescence.

|   | Cu:In | Zn:Cu | Name |
|---|---|---|---|
| A | 2.41 ± 0.03 |  | $CuIn_{0.4}S_2$ |
| B | 0.28 ± 0.03 |  | $Cu_{0.3}InS_2$ |
| C | 1.25 ± 0.05 | 0.216 ± 0.001 | $Cu(Zn)In_{0.8}S_2$ |
| D | 0.299 ± 0.009 | 2.810 ± 0.006 | $Cu_{0.3}(Zn)InS_2$ |
| E | 0.245 ± 0.005 | 10.41 ± 0.01 | $Cu_{0.2}(Zn)InS_2/ZnS$ |

The characterization results for $CuIn_{0.4}S_2$, $Cu_{0.3}InS_2$ and $Cu_{0.2}(Zn)InS_2/ZnS$ were previously published.[1]

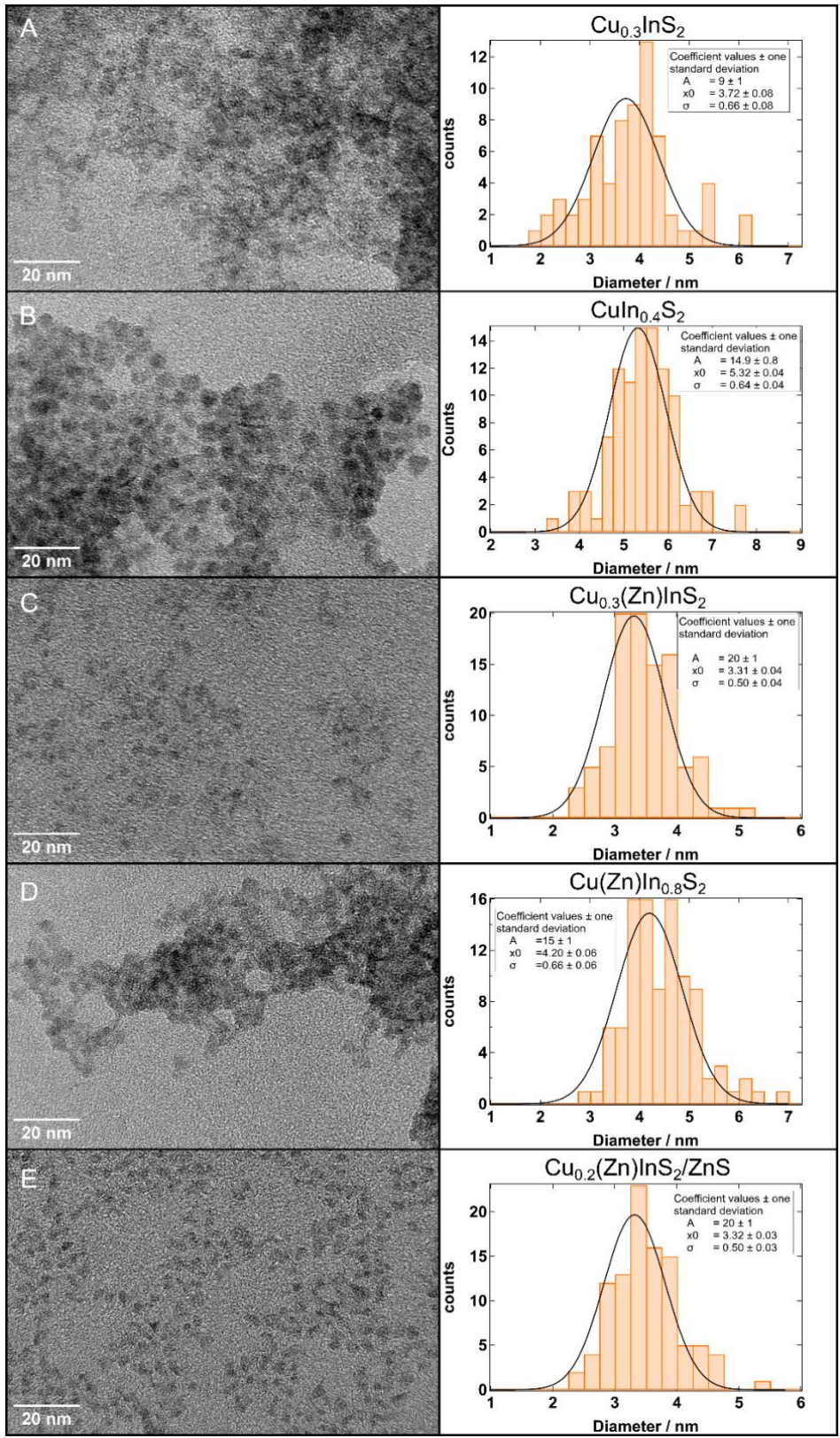

**Figure S1.** HR-TEM micrograph (left), and corresponding histograms (right) for the five samples under study. The histograms have been obtained with these and supplementary micrographs.

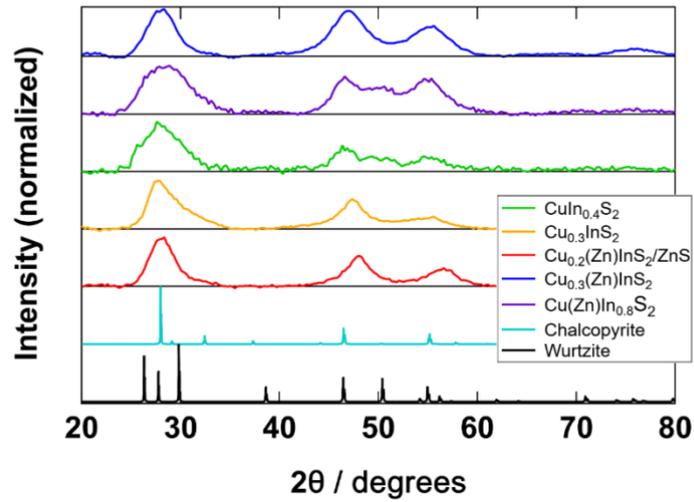

**Figure S2.** XRD patterns from the five samples and two reference spectra, chalcopyrite[2] and wurtzite,[3] calculated with VESTA.[4]

## 2. Complementary steady-state photoluminescence and UV-Vis absorption spectra

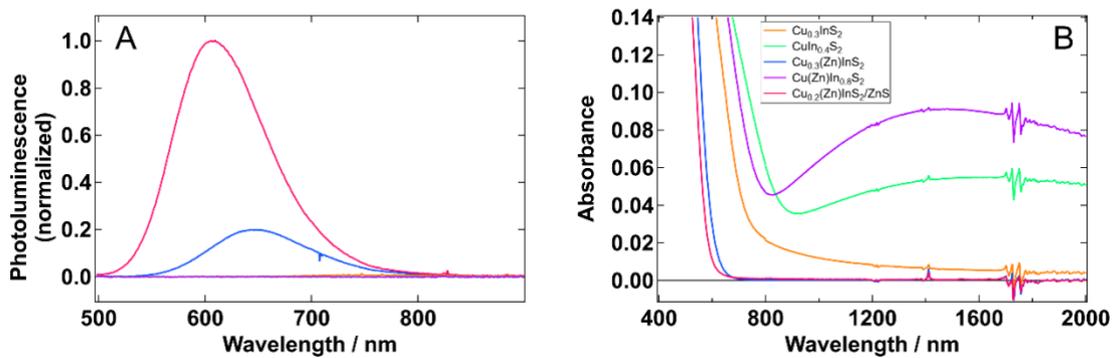

**Figure S3:** Linear photoluminescence spectra (A) and absorbance spectra (B) of the five samples extended to the Near infrared, where the stoichiometric samples show what can be assigned to a Localized surface plasmon resonance (LSPR)

## 3. Cu K-edge XANES and EXAFS

Cu K-edge XAS measurements were carried out during the same beamtime as the Zn and S K-edge experiments shown in Fig. 2. The results for $Cu_{0.3}InS_2$, $CuIn_{0.4}S_2$ and $Cu_{0.2}(Zn)InS_2/ZnS$ were already published in our previous article.[1]

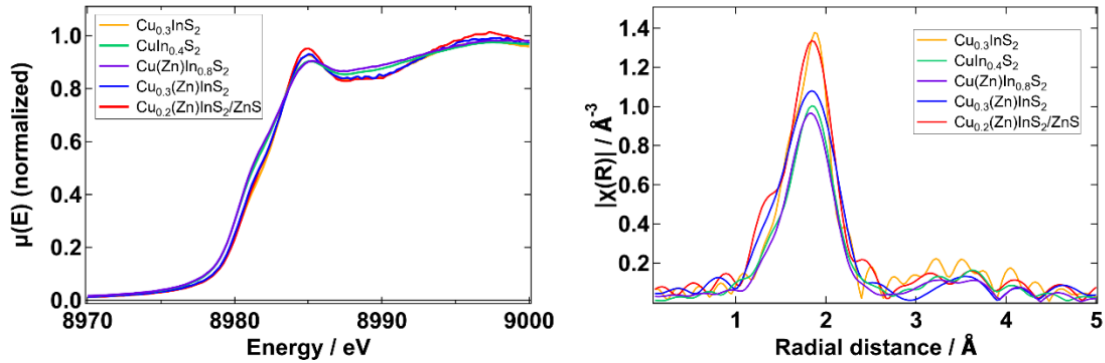

**Figure S4:** XANES (left) and EXAFS (right) spectra at the Cu K-edge for the five samples under study. Note the broader $1s \rightarrow 4p$ transition at 8984 eV for the stoichiometric samples, characteristic of a more disordered structure.

### 3.1 FEFF fitting results of Zn K-edge EXAFS

EXAFS analysis was carried out with the Artemis program,[5] applying the in-built FEFF6 for the fitting procedures. All samples were fitted in identical spectral ranges within the following limits:

k-range = 2.1 – 13 Å$^{-1}$
k-weight = 1
R-range = 1.25 – 3 Å

The background removal was carried out with k$^2$ weighting, a $R_{bkg}$ of 1.2 Å and values of $E_0$ of 9661.2 and 9661.8 eV.

The crystal structure of ZnS was obtained from a reference and used for the FEFF calculation.[6] The resulting single scattering paths involving S atoms (1$^{st}$ coordination shell) were used to fit the experimental data according to EXAFS equation:[7]

$$\chi(k) = \sum_j \frac{S_0^2 N_j f_j(k) e^{-2R_j/\lambda(k)} e^{-2k^2\sigma_j^2}}{kR_j^2} \sin[2kR_j + \delta_j(k)], \qquad (S1)$$

where our aim was to obtain statistically reliable/meaningful fitting of $N_j$ (coordination numbers), $R_j$, (first shell near neighbor distances,) and $\sigma_j$ (mean square disorder or Debye-Waller factors) parameters. The values of $S_0^2$ can be obtained from the literature or from the fit of a well-known reference.

In our case we set the value of $S_0^2 = 0.7$ as a reasonable estimation obtained from an EXAFS fit of bulk ZnS reference sample data (see Table 1). A k weight of 1 was used for the fit in order to maximize the first shell S scattering signal. The best-fit results are summarized in Table S2 and depicted in Fig. S2.

**Table S2:** Zn K-edge EXAFS FEFF fitting parameters for the single scattering path with the first shell of S atoms to a Zincblende ZnS structure.

| Structure | Sample | $N \cdot S_0^2$ | N | $\sigma^2$ | R |
|---|---|---|---|---|---|
| ZnS Zincblende | $Cu(Zn)In_{0.8}S_2$ | $3.0 \pm 0.4$ | $4.3 \pm 0.6$ | $0.007 \pm 0.002$ | $2.30 \pm 0.02$ |
| | $Cu_{0.3}(Zn)InS_2$ | $2.5 \pm 0.3$ | $3.5 \pm 0.5$ | $0.005 \pm 0.002$ | $2.33 \pm 0.01$ |
| | $Cu_{0.2}(Zn)InS_2/ZnS$ | $1.9 \pm 0.3$ | $2.6 \pm 0.4$ | $0.004 \pm 0.003$ | $2.34 \pm 0.02$ |

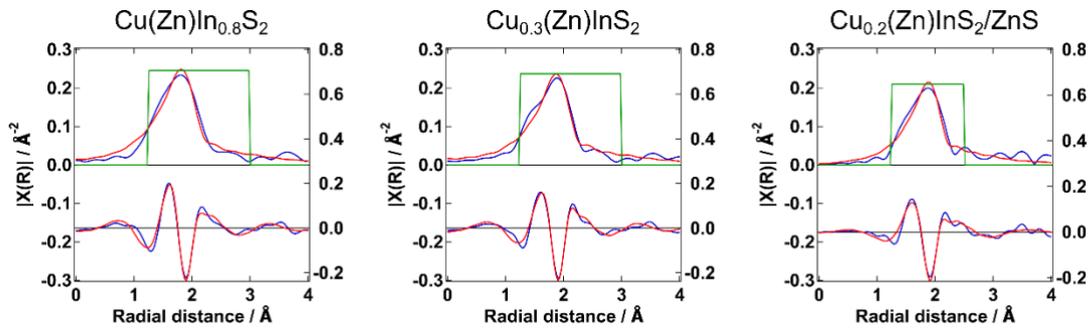

**Figure S5:** FEFF6 fits (red) for all CZIS QD samples (blue) at the Zn K-edge. The fits are carried out for a Zincblende ZnS structure. The actual disorder and mixture of structures in which Zn is located prevents us from obtaining excellent fits. However, it allows us a rough comparison between samples.

### 3.2 Linear combination fitting of XANES spectra

The fitting of XANES spectra using a linear superposition of different QD composition data were performed with the Athena program of the Demeter package.[5] In Fig. S3 we show the results of such fitting using a linear combination of $Cu_{0.2}(Zn)InS_2/ZnS$ and $CuIn_{0.4}S_2$ spectra.

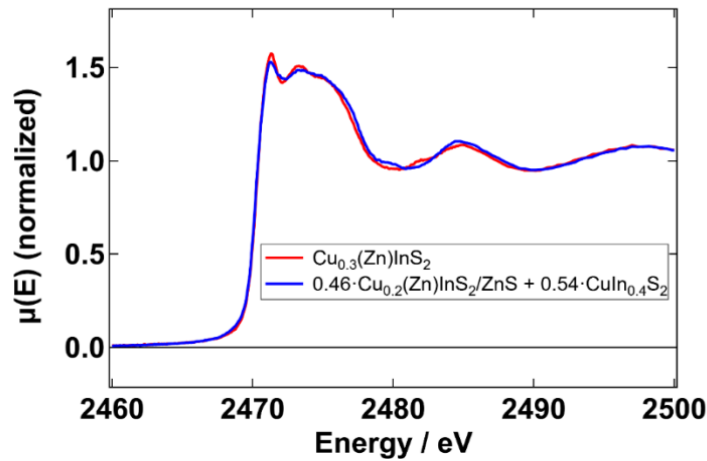

**Figure S6:** Linear combination fitting of $Cu_{0.3}(Zn)InS_2$ with $Cu_{0.2}(Zn)InS_2/ZnS$ and $CuIn_{0.4}S_2$.

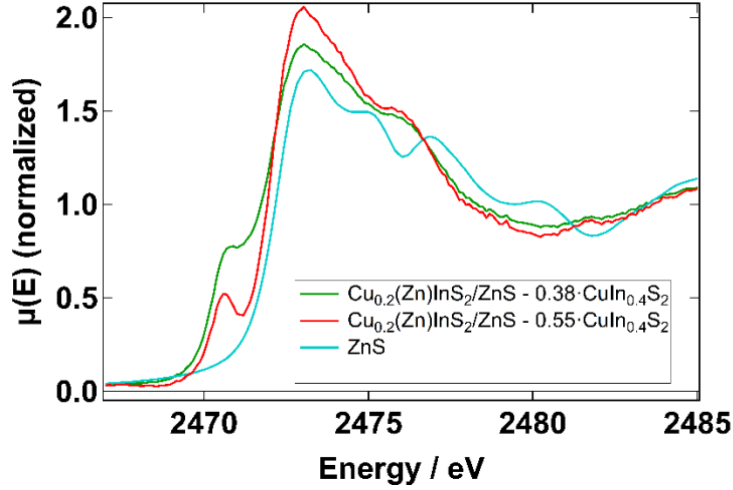

**Figure S7:** Extraction of the ZnS shell Sulfur K-edge spectrum. Due to the slight differences between samples, it is not possible to completely remove the 2471 eV peak. In addition, the bulk ZnS spectrum shows additional oscillations. We propose that they originate from either multiple-scattering or second shell scattering that is not significant in the thin and disordered ZnS shell of the quantum dots.

### 3.3 Shell thickness estimation

If we consider the crystalline structures of Fig. 1.B, we can make an estimation of the shell thickness for $Cu_{0.2}(Zn)InS_2/ZnS$ and $Cu_{0.3}(Zn)InS_2$.

Let us assume spherical QDs. The radius (r) of the QD is given by the TEM measurement.

Considering $V = \frac{4}{3}\pi r^3$, we can obtain the Volume (V) from the radius or viceversa.

From to the crystal structures of chalchopyrite[2] ($CuInS_2$) and zincblende[6] (ZnS) we can obtain the unit cell volume and the volume per S atom, obtaining 42.08 and 39.68 Å$^3$/S atom, respectively. Since both values are very close, we will consider them equal and directly translate proportion of S in each environment to volume.

With the proportion of S on the core and the shell we can estimate the radius of the core and the thickness of the shell (d). Based on Figs. S6 and S7 we consider proportions of 60% and 30% of ZnS for $Cu_{0.2}(Zn)InS_2/ZnS$ and $Cu_{0.3}(Zn)InS_2$. With this, we calculate the volume of the core, its radius and the shell thickness.

**Table S3:** Shell thickness calculations based on the particle size obtained with TEM and the S proportions obtained from XAS.

| QD | $R_{tot}$ / nm | V / nm$^3$ | $V_{core}$ / nm$^3$ | $r_{core}$ / nm | d / nm |
|---|---|---|---|---|---|
| $Cu_{0.2}(Zn)InS_2/ZnS$ | 1.66 | 19.1 | 7.66 | 1.22 | 0.44 |
| $Cu_{0.3}(Zn)InS_2$ | 1.66 | 19.0 | 13.3 | 1.47 | 0.19 |

For $Cu_{0.2}(Zn)InS_2/ZnS$ we obtain a shell thickness of 0.44 nm, close to the unit cell length of 0.54 nm, explaining the passivation. Meanwhile $Cu_{0.3}(Zn)InS_2$ has a shell thickness of 0.19 nm, considerably smaller than the ZnS unit cell. This would correspond to a monolayer of Zn atoms on the surface, giving higher chances of charge transfer through the shell.

## 3.4 Wavelet transform analysis

Wavelet transform (WT) analysis was carried out using the XWave code,[8] employing a Morlet wavelet. The parameters used were:

Resolution = 0.01

η = 7.5

σ = 0.5

k-weight = 3

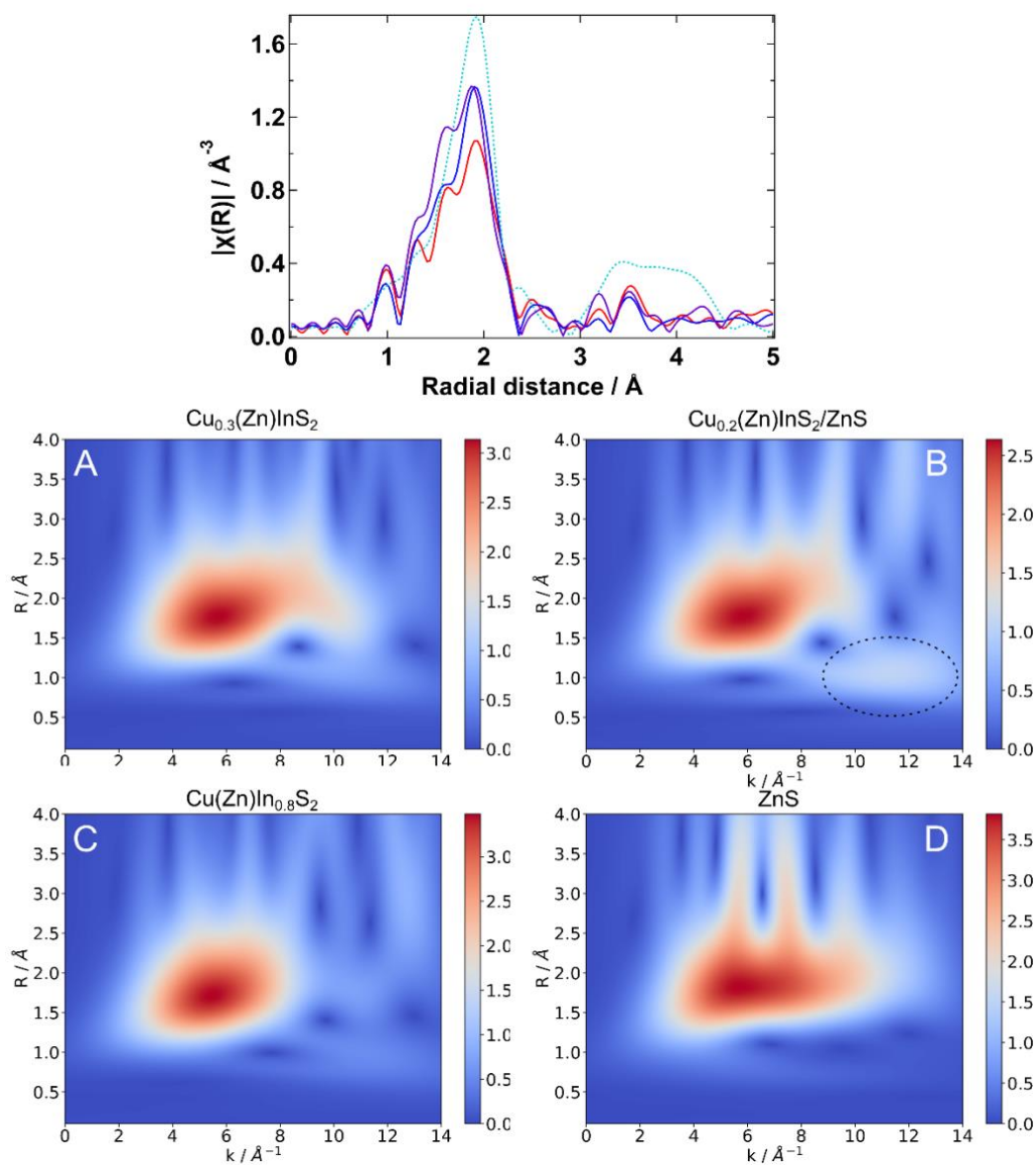

**Figure S8:** Alternative data treatment for the Zn K-edge EXAFS and wavelet calculation. The plots shown in Figs. 2 and 3 have been obtained applying a Rbkg value in Athena of 1.2, while these are obtained with a value of 0.95. This avoids suppressing the signal at around 1 Å, but introduces an artifact in the form of an oscillation across the spectrum.

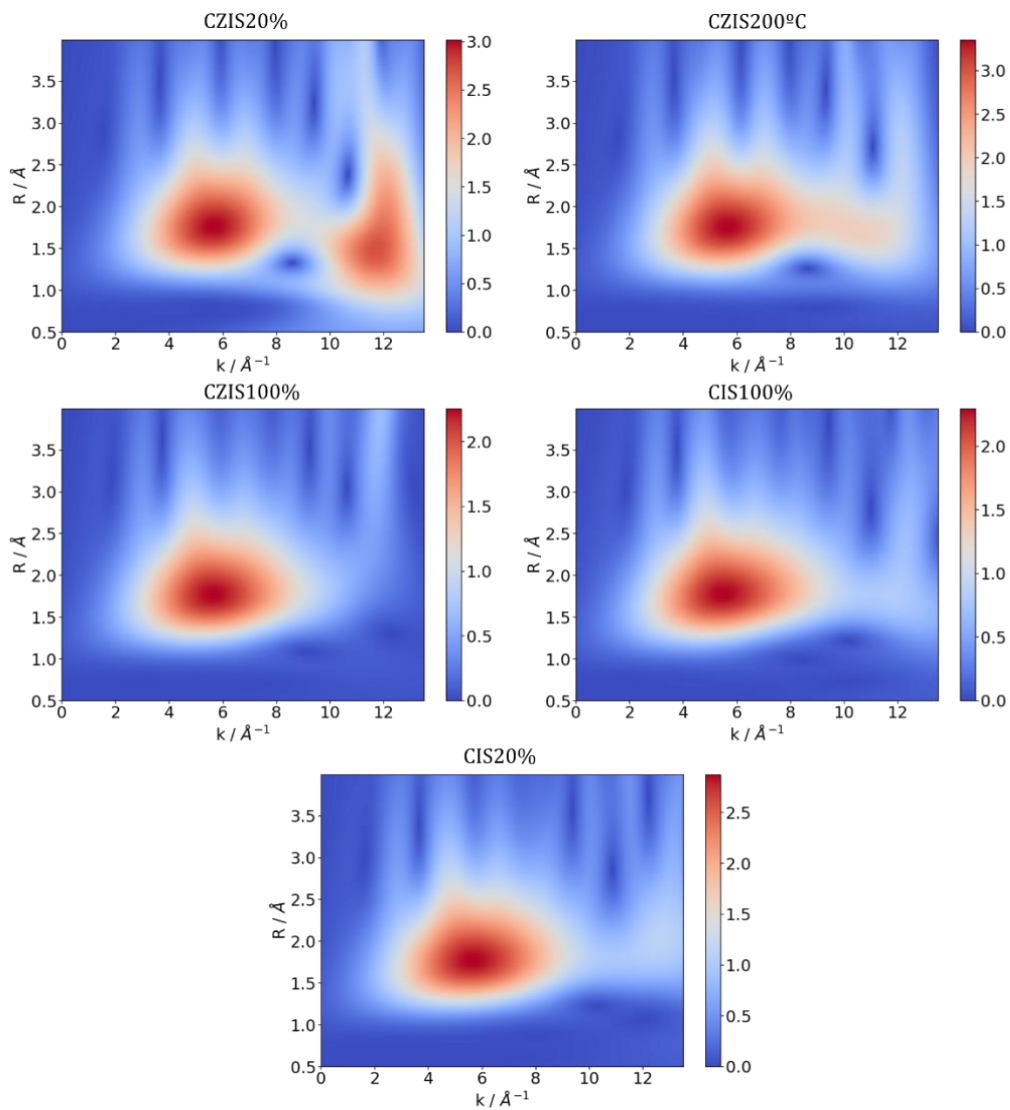

**Figure S9:** Wavelet transform of the Cu K-edge EXAFS of $Cu_{0.3}(Zn)InS_2$, $Cu_{0.2}(Zn)InS_2/ZnS$, $Cu(Zn)In_{0.8}S_2$, $CuIn_{0.4}S_2$, and $Cu_{0.3}InS_2$. We employed cubic k-weighting to maximize the contrast in the high k region of the spectra, the same as in case of Zn K-edge (Fig. 3 of the main text).

## 4. Time-resolved photoluminescence analysis

The time-resolved photoluminescence (TRPL) can also be fitted to a sum of exponential decays convoluted with Gaussian functions:

$$S(t) = A\frac{1}{2}\exp\left(\frac{4ln(2)w^2}{\tau^2} - \frac{t-t_0}{\tau}\right)\left(1 + \text{erf}\left(\frac{t-t_0}{4\sqrt{ln(2)}\,w} - \frac{2\sqrt{ln(2)}\,w}{\tau}\right)\right), \quad (S2)$$

where $w$ is the full with at half maximum (FWHM), $t_0$ is the time zero, $A$ is the amplitude and $\tau$ is the lifetime.

From the fitting of TRPL data we obtain intensity-weighted averages of the lifetimes from a penta-exponential decay fit and compare them to obtain the quenching efficiency ($\Phi_q$).

The results (Table S2) indicate lifetime quenching via hole transfer process of less than 10% for the most passivated sample ($Cu_{0.2}(Zn)InS_2/ZnS$), while it goes up to about 20% for $Cu_{0.3}(Zn)InS_2$. However, the multiexponential nature of the measured decays in these samples contributes to a larger uncertainty, compared to other systems, explaining the discrepancy with the time-integrated method showed in the main text. Nonetheless, comparing the decay traces in Fig. 4, we clearly observe that $Cu_{0.3}(Zn)InS_2$ is the most affected one, with the effect becoming present already at ultrafast timescales.

**Table S4**. Quenching results from intensity-averaged multiexponential fits.

| QD | QD-PTZ | | | QD-BQ | |
|---|---|---|---|---|---|
| TRPL | $\tau_{avg}$ / ps | $\tau_{avg}$ / ps | $\Phi_q$ | $\tau_{avg}$ / ps | $\Phi_q$ |
| $Cu_{0.2}(Zn)InS_2/ZnS$ | 288 ± 52 | 260 ± 160 | 0.09 | 14.3 ± 0.8 | 0.957 |
| $Cu_{0.3}(Zn)InS_2$ | 220 ± 20 | 174 ± 15 | 0.217 | | |

## 5. Complementary PL spectra

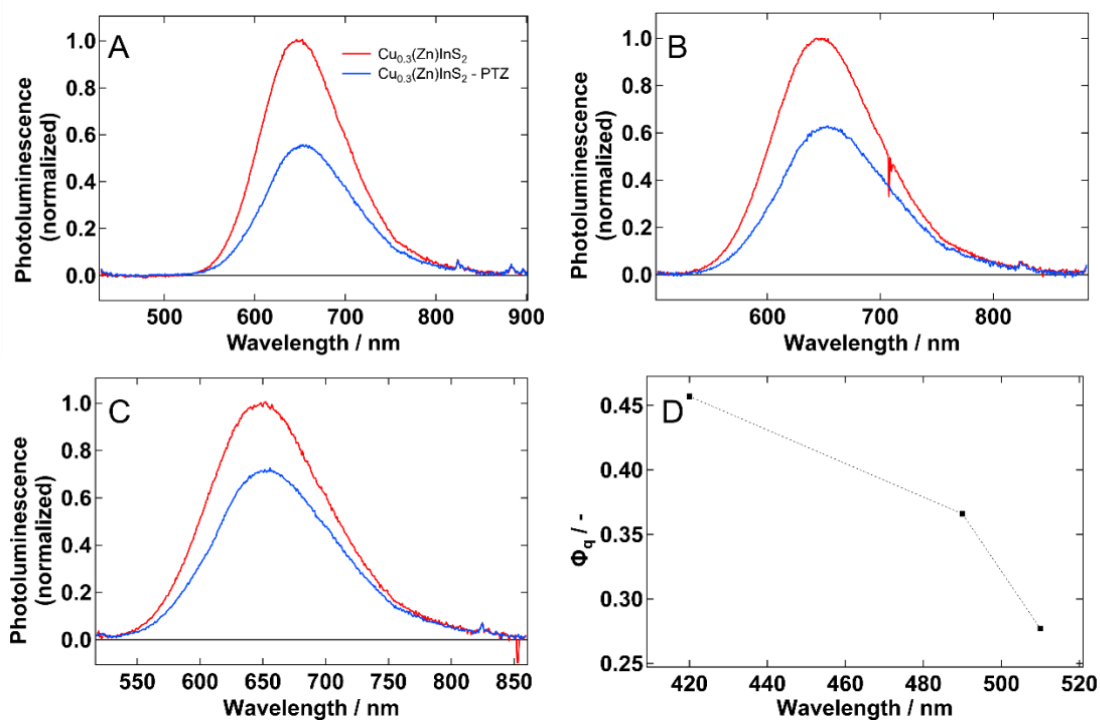

**Figure S10:** PL measurements of the quenching effects of PTZ where the excitation wavelength is A) 420 nm, B) 490 nm, and C) 510 nm. D) Wavelength-dependent quenching efficiency ($\Phi_q$).

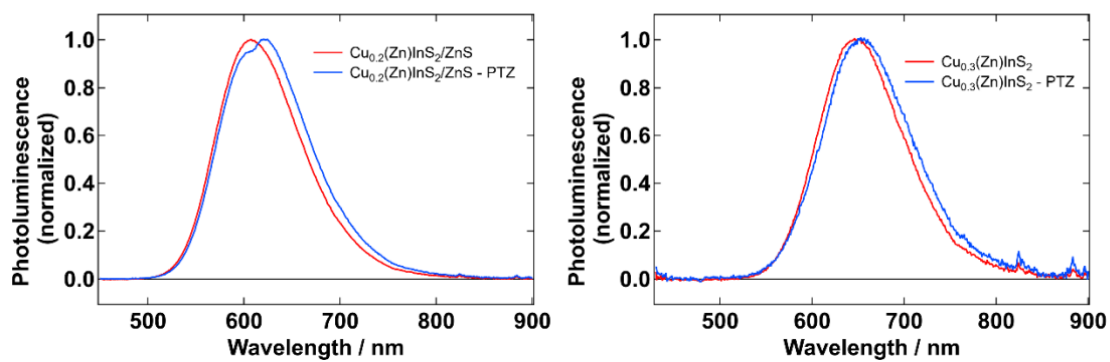

**Figure S11:** Normalized plots of $Cu_{0.2}(Zn)InS_2/ZnS$ (left) and $Cu_{0.3}(Zn)InS_2$ (right) with and without PTZ. The incorporation of PTZ slightly shifts the emission spectrum. CHS states located closer to the valence band will be more likely to finally transfer to PTZ, explaining the shift in the spectrum.

# References

(1) Burgos-Caminal, A.; Vale, B. R. C.; Fonseca, A. F. V.; Collet, E. P. P.; Hidalgo, J. F.; García, L.; Watson, L.; Borrell-Grueiro, O.; Corrales, M. E.; Choi, T.-K.; Katayama, T.; Fan, D.; Vega-Mayoral, V.; García-Orrit, S.; Nozawa, S.; Penfold, T. J.; Cabanillas-Gonzalez, J.; Adachi, S.-I.; Bañares, L.; Nogueira, A. F.; Padilha, L. A.; Schiavon, M. A.; Gawelda, W. Selective Tracking of Charge Carrier Dynamics in CuInS2 Quantum Dots. arXiv December 19, 2024. https://doi.org/10.48550/arXiv.2412.15418.

(2) Hahn, H.; Frank, G.; Klingler, W.; Meyer, A.-D.; Störger, G. Untersuchungen über ternäre Chalkogenide. V. Über einige ternäre Chalkogenide mit Chalkopyritstruktur. *Z. Für Anorg. Allg. Chem.* **1953**, *271* (3–4), 153–170. https://doi.org/10.1002/zaac.19532710307.

(3) Li, Q.; Zhai, L.; Zou, C.; Huang, X.; Zhang, L.; Yang, Y.; Chen, X.; Huang, S. Wurtzite CuInS2 and CuInxGa1−xS2 Nanoribbons: Synthesis, Optical and Photoelectrical Properties. *Nanoscale* **2013**, *5* (4), 1638–1648. https://doi.org/10.1039/C2NR33173J.

(4) Momma, K.; Izumi, F. Vesta 3 for Three-Dimensional Visualization of Crystal, Volumetric and Morphology Data. *J. Appl. Crystallogr.* **2011**, *44* (6), 1272–1276. https://doi.org/10.1107/S0021889811038970.

(5) Ravel, B.; Newville, M. ATHENA, ARTEMIS, HEPHAESTUS: Data Analysis for X-Ray Absorption Spectroscopy Using IFEFFIT. *J. Synchrotron Radiat.* **2005**, *12* (4), 537–541. https://doi.org/10.1107/S0909049505012719.

(6) Jumpertz, E. A. Electron-Density Distribution in Zinc Blende. *Z. Für Elektrochem. Angew. Phys. Chem.* **1955**, No. 59, 419–425.

(7) Newville, M. EXAFS Analysis with Feff, Larch, Artemis, 2018. https://millenia.cars.aps.anl.gov/videos/FundamentalsOfXAFS/UsingFeff.pdf.

(8) Penfold, T. J.; Tavernelli, I.; Milne, C. J.; Reinhard, M.; Nahhas, A. E.; Abela, R.; Rothlisberger, U.; Chergui, M. A Wavelet Analysis for the X-Ray Absorption Spectra of Molecules. *J. Chem. Phys.* **2013**, *138* (1), 014104. https://doi.org/10.1063/1.4772766.